\documentclass[letterpaper,journal]{IEEEtran}
\usepackage{amsmath,amsfonts}
\usepackage[ruled,vlined,linesnumbered]{algorithm2e}
\usepackage{algorithmic}
\usepackage{array}
\usepackage{textcomp}
\usepackage{stfloats}
\usepackage{url}
\usepackage{verbatim}
\usepackage{graphicx}
\usepackage{rotating}
\usepackage{tikz}
\usetikzlibrary{arrows.meta,positioning,fit,backgrounds,decorations.pathreplacing}
\usepackage{cite}

\usepackage{amssymb}
\usepackage{booktabs}
\usepackage{bm}
\usepackage{bbm}
\usepackage{mathrsfs}
\usepackage{mathtools}
\usepackage{tabularx}

\usepackage{amsthm}
\usepackage{thmtools}
\usepackage{thm-restate}

\newtheorem{theorem}{Theorem}  
\newtheorem{proposition}{Proposition}  

\usepackage{cuted}
\usepackage{siunitx}
\usepackage{placeins}
\usepackage{multirow}
\usepackage{nicefrac}
\usepackage{orcidlink}
\usepackage{hyperref}

\newcommand{\e}[1]{\times 10^{#1}}
\newcommand{\comp}[1]{\overline{#1}}          

\newcommand{\xcoord}[2][]{\ensuremath{\ifx\relax#1\relax X[#2]\else #1[#2]\fi}}
\newcommand{\xkeep}[2][]{\ensuremath{\ifx\relax#1\relax X_{#2}\else \bigl(#1\bigr)_{#2}\fi}}
\newcommand{\xprune}[2][]{\ensuremath{\ifx\relax#1\relax X_{\comp{#2}}\else \bigl(#1\bigr)_{\comp{#2}}\fi}}

\begin{document}

\title{CCMP: Class-Conditional Multicollinearity Pruning for High-Dimensional Feature Selection}

\author{Abdullah Al Bashit\,\orcidlink{0000-0002-7726-8916},
        Prakash Nepal\,\orcidlink{0000-0003-2059-0426}, and
        Lee Makowski\,\orcidlink{0000-0003-0202-9264}
    \thanks{This work was supported in part by the National Institutes of Health (R21-AG068972; RF1AG079946). \textit{(Corresponding author: Abdullah Al Bashit.)}}
\thanks{The authors are with the Department of Bioengineering, Northeastern University, Boston, MA, USA (e-mail: a.bashit@northeastern.edu; p.nepal@northeastern.edu; l.makowski@northeastern.edu).}
\thanks{This article has supplementary downloadable material available at \nolinkurl{https://github.com/abdullah-al-bashit/CCMP}, provided by the authors.}
}

\markboth{IEEE Transactions on Signal Processing,~2026}%
{Bashit \MakeLowercase{\textit{et al.}}: Class-Conditional Multicollinearity Pruning for Feature Selection}


\maketitle
\begin{abstract}
In high-dimensional scientific measurements, many features in a single sample often capture the same underlying physical attributes, producing severe multicollinearity. Supervised feature selectors fit on the entire dataset, so a feature can be informative for classification overall while still redundant with another feature in the same class. Class-conditional selectors, by contrast, trade some accuracy for a narrower, per-class view, since they never see data outside one class, but this lets new classes be added without refitting the whole model. They still cannot determine, however, which feature to drop from a correlated group. To close this gap, we propose \emph{class-conditional multicollinearity pruning} (CCMP), which iteratively prunes one feature from every highly correlated pair, per class. Four theoretical guarantees establish explicit bounds on the error induced by pruning under stated assumptions. Across four benchmark datasets spanning spectroscopic and hyperspectral imaging modalities, CCMP retains as few as 42 of 1531 features (3\%) and as many as 60 of 145 (41\%), outperforms every class-conditional baseline, and remains competitive with supervised methods.
\end{abstract}

\begin{IEEEkeywords}
Multicollinearity, class conditional pruning, correlation-guided feature pruning, feature selection, feature pruning.
\end{IEEEkeywords}

\section{Introduction}

\IEEEPARstart{H}{igh-dimensional} measurements typically collected from spectroscopy, hyperspectral sensing, and other signal-processing domains contain hundreds to thousands of features on a fine, ordered grid. Because many grid points across the sample reflect the same underlying physical attributes, features show strong correlation, producing severe \emph{multicollinearity}: a near-linear dependence among predictors that collapses the effective rank of the feature space~\cite{belsley1980regression, montgomery2012introduction}. Multicollinearity inflates estimator variance, destabilizes predictions, and makes it difficult to discern which individual features carry information that actually helps separate the classes~\cite{Hastie2009}. The problem is compounded in the limited-sample setting common to scientific measurements, where the number of observations $N$ is comparable to, or smaller than, the feature dimension $d$, so that systematic dimensionality reduction is needed before any classifier can generalize to unseen samples.

\begin{figure}[t]
\centering
\resizebox{\linewidth}{!}{%
\begin{tikzpicture}[
  b/.style={draw, rounded corners=2pt, minimum height=6mm, align=center, font=\footnotesize},
  arr/.style={-{Stealth[scale=0.8]}, thick},
]
  \node[b, fill=gray!8,    minimum width=13mm]                     (data)  {$X_{\mathrm{tr}},\mathbf{y}$};
  \node[b, fill=blue!12,   minimum width=22mm, right=6mm of data]  (corr)  {per-class $|\boldsymbol{\mathcal{C}}^{(y)}|$};
  \node[b, fill=teal!10,   minimum width=16mm, right=6mm of corr]  (tau)   {\raisebox{-1.5ex}{$\tau$ sweep}};
  \node[b, fill=blue!12,   minimum width=20mm, right=6mm of tau]   (prune) {greedy prune\\per class};
  \node[b, fill=orange!14, minimum width=22mm, below=6mm of prune] (union) {union $\bigcup_y$};
  \node[b, fill=teal!10,   minimum width=18mm, left=6mm of union]  (aic)   {$\kappa(\tau)\to\tau^\star$};
  \node[b, fill=green!12,  minimum width=14mm, left=6mm of aic]    (Q)     {$\widehat{\mathcal{Q}}_{\tau^\star}$};
  \node[b, fill=blue!8,    minimum width=14mm, left=6mm of Q]      (clf)   {Classifier};
  \draw[arr] (data)  -- (corr);
  \draw[arr] (corr)  -- (tau);
  \draw[arr] (tau)   -- (prune);
  \draw[arr] (prune) -- (union);
  \draw[arr] (union) -- (aic) node[midway, above, font=\scriptsize, text=gray!70!black] {$\widehat{\mathcal{Q}}_\tau$};
  \draw[arr] (aic)   -- (Q);
  \draw[arr] (Q)     -- (clf);
  \node[font=\scriptsize, text=gray!70!black, below=0.5mm of aic] {\textit{elbow rule}};
\end{tikzpicture}%
}
\caption{Overview of CCMP. Training data $X_{\mathrm{tr}}$ with labels $\mathbf{y}$ produce
per-class absolute Pearson matrices $|\boldsymbol{\mathcal{C}}^{(y)}|$. For each candidate cutoff
$\tau$ in a grid search, greedy pruning runs per class, and the retained features are
unioned across classes into the kept feature set $\widehat{\mathcal{Q}}_\tau$. A geometric elbow rule on the Redundancy
Retention Score $\kappa(\tau)$ then selects $\tau^\star$, giving the final kept feature set $\widehat{\mathcal{Q}}_{\tau^\star}$
for the downstream classifier.}
\label{fig:overview_pipeline}
\end{figure}

Regularization addresses variance inflation by adding a penalty term to a single model fit on the training set \((X_{\mathrm{tr}}, \mathbf{y})\) across all classes. Lasso~\cite{tibshirani1996regression} promotes sparsity, but when correlations computed over this data are very high among a group of features, it tends to keep only one feature from the group~\cite{zou2005regularization}. ElasticNet~\cite{zou2005regularization}, in contrast, groups correlated features so that sparsity no longer depends on an arbitrary choice. In both methods, correlation is measured across $X_{\mathrm{tr}}$. Thus, a large fitted coefficient shows how useful a feature is for separating classes overall, not whether it is redundant with another feature in the same class.

Ranking features by importance produces a discrete index set of retained features (we use \emph{retained} and \emph{kept} interchangeably). Supervised selectors, including feature rankings from Lasso and ElasticNet above, mutual information (Mut.\ Info.)~\cite{battiti1994using}, and \(\ell_1\)-penalized support vector machines (\(\ell_1\)-SVM)~\cite{fan2008liblinear}, likewise need access to data from every class. Class-conditional selectors restrict that data to the within-class distribution. Because class-conditional selectors never see the entire dataset, they also tend to underperform supervised selectors on classification accuracy. Still, a feature's score for one class does not change if other classes are added to or removed from the dataset, and computing it requires only that class's own samples. This makes class-conditional selectors particularly useful as a dataset expands to include new classes, considering that supervised selectors would otherwise need to be refit from scratch. Among class-conditional selectors, the Laplacian score (LapScore)~\cite{he2005laplacian} measures local smoothness across same-class neighbors, the median absolute deviation (MAD) filter ranks features by per-class spread, and ridge reconstruction (RidgeRecon)~\cite{hoerl1970ridge} scores each feature by how well it is linearly reconstructed from the rest within its class. None of these approaches, however, can determine which of two correlated features to drop. This paper closes that gap with \emph{class-conditional multicollinearity pruning} (CCMP), which prunes redundant features per class, outperforming other class-conditional methods and narrowing the accuracy gap to supervised selectors.

Fig.~\ref{fig:overview_pipeline} summarizes CCMP. Sweeping a grid of candidate correlation cutoffs $\tau$, greedy pruning per class discards one feature from every pair whose absolute correlation exceeds $\tau$, and the class-wise union of what remains gives the kept feature set $\widehat{\mathcal{Q}}_\tau$. A geometric elbow rule picks $\tau^\star$ by maximizing the Redundancy Retention Score $\kappa(\tau)$, which captures how well the kept set explains what was discarded, yielding the final kept feature set $\widehat{\mathcal{Q}}_{\tau^\star}$ for the downstream classifier. Beyond this algorithm, we establish formal guarantees on the error induced by pruning and validate the method against a comprehensive baseline suite on several high-dimensional, strongly correlated benchmarks. This paper presents the following contributions.

\begin{itemize}
    \item \textbf{Class-conditional multicollinearity pruning (CCMP).} Per class, CCMP (Algorithm~\ref{alg:class_conditional_pruning}) prunes highly correlated features using a correlation cutoff selected by a geometric elbow rule on the Redundancy Retention Score. As a result, the kept features are mutually uncorrelated per class and lie in the original physical measurement domain. An ablation study further confirms the robustness of this selection across downstream classifiers.

    \item \textbf{Theoretical guarantees.} Four formal results constrain the error induced by pruning under explicitly stated assumptions. These comprise a Bayes risk collapse bound (Proposition~\ref{proposition:bayes-risk-collapse}), a gradient covariance lower bound on irreducible error (Theorem~\ref{thm:psi-bound}), a H\"older smoothness curvature penalty (Corollary~\ref{cor:holder-gradient-bound}), and a tightness result in the affine regime (Proposition~\ref{prop:affine-segment}).

    \item \textbf{Experimental validation.} We evaluate CCMP against supervised selectors (Lasso, ElasticNet, \(\ell_1\)-SVM, Mut.\ Info.), class-conditional baselines (LapScore, MAD, RidgeRecon), and an unpruned full-feature baseline, on four high-dimensional, strongly correlated datasets. CCMP outperforms every class-conditional baseline, approaches supervised-selector performance, stays within 1.7\% of the full-feature baseline, and experimentally satisfies the theoretical bounds. 

\end{itemize}

\begin{figure*}[t]
\centering
\resizebox{0.98\textwidth}{!}{%
\input{fig/method_concept}
}
\caption{CCMP on a $d\!=\!4$ example.
(a) Per-class absolute correlation matrices. Orange cells exceed $\tau^\star$.
Sweeping $\tau$ tries each candidate against these matrices, with $\tau^\star\!=\!0.80$
highlighted as the value the elbow rule selects in panel (c).
(b) Greedy pruning discards one feature from each over-threshold pair, per class.
Class~1 drops $f_3$ and keeps $\{f_1,f_2,f_4\}$, while class~2 drops $f_3$ then $f_4$
and keeps $\{f_1,f_2\}$.
(c) The union of both classes' kept sets retains $\{f_1,f_2,f_4\}$. $f_3$ is dropped
by both classes and excluded. The elbow rule selects $\tau^\star$ from the $\kappa(\tau)$
curve, and the retained features are passed to the classifier, shown schematically
with an arbitrary decision boundary separating the two classes.}
\label{fig:method_concept}
\end{figure*}

\section{Proposed Method}
\label{sec:proposed_method}

This section formalizes CCMP, illustrated in Fig.~\ref{fig:method_concept} on a $d{=}4$ example. We define the class-conditional correlation matrices and describe how correlated pairs are discarded to produce the kept set $\widehat{\mathcal{Q}}_\tau$, then select the cutoff $\tau^\star$ via a geometric elbow rule on $\kappa(\tau)$. This section closes by deriving the algorithm's computational cost and stating the formal guarantees on the error induced by pruning.

\subsection{Class-Conditional Multicollinearity Pruning (CCMP)}
\label{sec:class-conditional-pruning}
We consider a labeled dataset with \(C\) classes and \(N\) samples, where \(X\in\mathbb{R}^d\) is a feature vector belonging to class \(y\in[C]:=\{1,\dots,C\}\). For each class \(y\), we compute a separate absolute Pearson correlation matrix \(\boldsymbol{\mathcal{C}}^{(y)}\):
\begin{equation}
\label{eq:class_conditional_abs_corr}
\boldsymbol{\mathcal{C}}^{(y)} := \bigl|\,\mathbf{D}_y^{-1/2}\,\boldsymbol{\Sigma}^{(y)}\,\mathbf{D}_y^{-1/2}\,\bigr|,
\qquad
\mathbf{D}_y := \operatorname{diag}\!\big(\boldsymbol{\Sigma}^{(y)}\big).
\end{equation}
The resulting \(\boldsymbol{\mathcal{C}}^{(y)}\in[0,1]^{d\times d}\) is invariant to per-feature shifts and scales.

We sweep a grid of candidate correlation cutoffs \(\tau\in\mathcal{T}\subset(0,1)\). For each \(\tau\), we define the correlated pair set \(\mathcal{P}_\tau^{(y)}\) as all strictly upper-triangular index pairs \((a,b)\) whose absolute correlation exceeds \(\tau\):
\begin{equation}
\label{eq:pair_set}
\mathcal{P}_\tau^{(y)} := \left\{\, (a,b)\in\{1,\dots,d\}^2 : a<b,\ \boldsymbol{\mathcal{C}}^{(y)}_{ab} > \tau \,\right\}.
\end{equation}

Pairs are sorted by descending absolute correlation and processed in that order. Each step greedily discards one feature from the current pair. A discard set \(\mathscr{R}^{(y)}\gets\varnothing\) accumulates removed features, and a pair \((a,b)\) is skipped once one of its features has already been discarded.

Let \(\mathcal{V}_{ab}^{(y)}\) denote the active set, every feature not yet discarded and not equal to \(a\) or \(b\),
\begin{equation}
\label{eq:valid_set_ab}
\mathcal{V}_{ab}^{(y)} := \left\{ j \in [d] \;\middle|\; j \notin \mathscr{R}^{(y)},\ j \ne a,\ j \ne b \right\}.
\end{equation}

We then compute the mean of $\boldsymbol{\mathcal{C}}^{(y)}_{a\ell}$ and $\boldsymbol{\mathcal{C}}^{(y)}_{b\ell}$ over the shared $\ell \in \mathcal{V}_{ab}^{(y)}$:
\begin{equation}
\label{eq:feature_connectivity}
\overline{c}_a \;:=\; \frac{1}{\bigl|\mathcal{V}_{ab}^{(y)}\bigr|}\!\sum_{\ell\in\mathcal{V}_{ab}^{(y)}} \boldsymbol{\mathcal{C}}^{(y)}_{a\ell},
\qquad
\overline{c}_b \;:=\; \frac{1}{\bigl|\mathcal{V}_{ab}^{(y)}\bigr|}\!\sum_{\ell\in\mathcal{V}_{ab}^{(y)}} \boldsymbol{\mathcal{C}}^{(y)}_{b\ell},
\end{equation}

The feature with higher mean absolute correlation ($\overline{c}_a$ or $\overline{c}_b$) is added to $\mathscr{R}^{(y)}$:
\begin{equation}
\label{eq:discard_update}
j^\star \;:=\; \arg\max\{\overline{c}_a,\overline{c}_b\},
\qquad
\mathscr{R}^{(y)} \;\gets\; \mathscr{R}^{(y)} \cup \{j^\star\}.
\end{equation}

Once all pairs in $\mathcal{P}_\tau^{(y)}$ have been processed, the retained set for class $y$ is $\widehat{\mathcal{Q}}_\tau^{(y)} := [d]\setminus \mathscr{R}^{(y)}$.

Their class-wise union gives \(\widehat{\mathcal{Q}}_\tau\) for this candidate \(\tau\),
\begin{equation}
\label{eq:pruned-set}
\widehat{\mathcal{Q}}_\tau \;:=\; \bigcup_{y\in[C]} \widehat{\mathcal{Q}}_\tau^{(y)}.
\end{equation}

\subsection{Pruning-Level Selection}
\label{sec:tau-selection}
Using the per-class kept sets \(\widehat{\mathcal{Q}}_\tau^{(y)}\) and discard sets \(\mathscr{R}^{(y)}\), we introduce the \emph{Redundancy Retention Score} (RRS) \(\kappa(\tau)\) to quantify how well the kept features represent the dropped ones. For each dropped feature \(i\in\mathscr{R}^{(y)}\), let
\begin{equation}
\label{eq:kappa-max}
m_i^{(y)}(\tau) \;:=\; \max_{j\in\widehat{\mathcal{Q}}_\tau^{(y)}} \boldsymbol{\mathcal{C}}^{(y)}_{ij}
\end{equation}
be the highest absolute correlation between feature \(i\) and any kept feature. The per-class score averages \(m_i^{(y)}(\tau)\) over all dropped features \(i\in\mathscr{R}^{(y)}\),
\begin{equation}
\label{eq:kappa-class}
\kappa^{(y)}(\tau) \;:=\;
\begin{cases}
1, & \mathscr{R}^{(y)} = \varnothing, \\[4pt]
\dfrac{1}{|\mathscr{R}^{(y)}|}\displaystyle\sum_{i\in\mathscr{R}^{(y)}} m_i^{(y)}(\tau), & \text{otherwise},
\end{cases}
\end{equation}
and the RRS is
\begin{equation}
\label{eq:kappa-def}
\kappa(\tau) \;:=\; \frac{1}{C}\sum_{y\in[C]} \kappa^{(y)}(\tau).
\end{equation}
A large \(\kappa(\tau)\) means pruned features are, on average, well represented by some retained feature. \(\kappa(\tau)\in(0,1]\), equal to \(1\) when nothing is dropped.

We select \(\tau\) with a geometric elbow rule on the curve \(\{(k_t,\kappa_t)\}_{t=1}^{T}\) traced over the candidate grid \(\mathcal{T}\), where \(k_t := |\widehat{\mathcal{Q}}_{\tau_t}|\) (\(\widehat{\mathcal{Q}}_{\tau_t} := \bigcup_y\widehat{\mathcal{Q}}_{\tau_t}^{(y)}\)) and \(\kappa_t := \kappa(\tau_t)\). Writing \(p_t := (\tilde k_t, \tilde\kappa_t)\) for the min-max normalized point (each component rescaled to \([0,1]\) over \(t\)), and \(a := p_1\), \(b := p_T\), \(\hat u := (b-a)/\|b-a\|\), the elbow point maximizes the perpendicular distance from the chord connecting the grid's endpoints,
\begin{equation}
\label{eq:tau-star}
\tau^\star \;:=\; \arg\max_{t}\, \bigl\|\,(p_t-a) - \bigl[(p_t-a)^\top\hat u\bigr]\hat u \,\bigr\|, \qquad \widehat{\mathcal{Q}} := \widehat{\mathcal{Q}}_{\tau^\star}.
\end{equation}

This rule requires no distributional assumption or sample-size correction, and identifies the point of steepest diminishing return in the compression-versus-redundancy-retention trade-off. Algorithm~\ref{alg:class_conditional_pruning} details the full procedure in pseudocode.

\begin{algorithm}[!t]
\caption{Class-Conditional Multicollinearity Pruning (CCMP)}
\label{alg:class_conditional_pruning}
\SetKwInOut{Require}{Require}
\SetKwInOut{Ensure}{Ensure}
\Require{Training data \((X_{\mathrm{tr}},\mathbf{y})\), \(X_{\mathrm{tr}}\in\mathbb{R}^{N\times d}\); candidate grid \(\mathcal{T}\subset(0,1)\)}
\Ensure{Retained index set \(\widehat{\mathcal{Q}}_{\tau^\star}\)}
\BlankLine
\For{each class \(y\in[C]\)}{
  \(X^{(y)} \gets \{x_i : y_i = y\}\)\;
  Compute \(\boldsymbol{\mathcal{C}}^{(y)}\) via \eqref{eq:class_conditional_abs_corr} and cache\tcp*{Done once per class}
}
\For{each \(\tau\in\mathcal{T}\)}{
  \(\widehat{\mathcal{Q}}_\tau \gets \varnothing\)\;
  \For{each class \(y\in[C]\)}{
    \eIf{\(|X^{(y)}| < 2\)}{
      \(\widehat{\mathcal{Q}}_\tau^{(y)} \gets [d]\)\tcp*{Too few samples, retain all}
    }{
      Form \(\mathcal{P}_\tau^{(y)}\) via \eqref{eq:pair_set}, sorted descending by \(\boldsymbol{\mathcal{C}}^{(y)}_{ab}\)\;
      \(\mathscr{R}^{(y)} \gets \varnothing\)\;
      \For{each \((a,b)\in\mathcal{P}_\tau^{(y)}\)}{
        \If{\(a\in\mathscr{R}^{(y)}\) \textbf{or} \(b\in\mathscr{R}^{(y)}\)}{
          \textbf{continue}\tcp*{Skip: one already removed}
        }
        Form \(\mathcal{V}_{ab}^{(y)}\) via \eqref{eq:valid_set_ab}. Compute \(\overline{c}_a\), \(\overline{c}_b\) via \eqref{eq:feature_connectivity}\;
        Update \(j^\star\) and \(\mathscr{R}^{(y)}\) via \eqref{eq:discard_update}\;
      }
      \(\widehat{\mathcal{Q}}_\tau^{(y)} \gets [d]\setminus\mathscr{R}^{(y)}\)\;
    }
    Update \(\widehat{\mathcal{Q}}_\tau\) via \eqref{eq:pruned-set}\;
  }
  Compute \(\kappa(\tau)\) via \eqref{eq:kappa-def}\;
}
Select \(\tau^\star\) and \(\widehat{\mathcal{Q}}_{\tau^\star}\) via \eqref{eq:tau-star}\;
\KwRet{\(\widehat{\mathcal{Q}}_{\tau^\star}\)}
\end{algorithm}

\subsection{Computational Complexity}
\label{sec:complexity}
The computational cost of CCMP breaks into two stages, correlation-matrix construction and per-cutoff pruning. The \(C\) per-class correlation matrices are built at \(\mathcal{O}(Nd^2)\) and cached across the sweep. For each of the \(T\) candidate cutoffs, the greedy pair-removal loop costs \(\mathcal{O}(Cd^3)\). The Redundancy Retention Score \(\kappa(\tau)\) reuses this same cache, costing \(\mathcal{O}(Cd^2)\) per cutoff. This is negligible next to the pruning loop's \(\mathcal{O}(Cd^3)\). The total is
\[
  \mathcal{O}\!\bigl(Nd^2 + T\,Cd^3\bigr).
\]
Because \(T\) is a small fixed grid, the one-time \(\mathcal{O}(Nd^2)\) correlation step dominates whenever \(N \gtrsim TCd\). After \(\tau^\star\) is fixed, all downstream inference runs on \(k\) features, reducing classifier cost by a factor of \(d/k\).

\subsection{Theoretical Guarantees}
\label{subsec:theoretical-guarantees}

Correlation-based pruning reduces redundancy, but it can remove class-discriminative features, raising predictive risk~\cite{Hastie2009}. We formalize this with four guarantees, stated under explicit assumptions. First, a Bayes risk collapse when the informative subset is removed. Second, a gradient covariance lower bound on the irreducible error induced by pruning. Third, a H\"older smoothness corollary that converts the bound into a closed form penalty. Fourth, a tightness guarantee in the affine regime. Collectively, they characterize when pruning is reliable versus brittle and show how risk scales with the dependence structure (pairwise correlation and class conditional covariance) and with model smoothness (the H\"older exponent \(\alpha\)).

The Bayes risk collapse is conditional on a sufficiency assumption.
\begin{quote}
\textbf{Assumption~1 (Sufficiency).} Let \( (\Omega, \mathcal{F}, \mathbb{P}) \) be a probability space, where \( X : \Omega \to \mathbb{R}^d \) is a random vector and \( Y : \Omega \to [C] := \{1, \dots, C\} \) is a discrete random variable. There exists an informative index set \(\mathcal{Q}_\tau^\star\subseteq[d]\) such that \(\mathbb{P}(Y\mid X)=\mathbb{P}(Y\mid X_{\mathcal{Q}_\tau^\star})\), equivalently \(Y\perp X_{[d]\setminus\mathcal{Q}_\tau^\star}\mid X_{\mathcal{Q}_\tau^\star}\).
\end{quote}

\begin{proposition}[Information loss risk]
\label{proposition:bayes-risk-collapse}
Under Assumption~1, with uninformative indices \( \overline{\mathcal{Q}_\tau^\star} := [d] \setminus \mathcal{Q}_\tau^\star \), the Bayes risk over the remaining features \( X_{\overline{\mathcal{Q}_\tau^\star}}, \) defined as the infimum over all measurable functions $\phi: \mathbb{R}^{|\overline{\mathcal{Q}_\tau^\star}|} \to [C]$, collapses to the maximum class prior:
\begin{equation}
    \inf_{\phi: \mathbb{R}^{|\overline{\mathcal{Q}_\tau^\star}|} \to [C]} \mathbb{P}\left(\phi(X_{\overline{\mathcal{Q}_\tau^\star}}) \ne Y\right)
= 1 - \max_{c \in [C]} \mathbb{P}(Y = c).
\end{equation}
\end{proposition}

\paragraph*{Proof} See Appendix~\ref{app:prop-bayes}.

Under Assumption~1, if pruning removes the informative indices \(\mathcal{Q}_\tau^\star\), then no classifier built on the remaining indices \(\overline{\mathcal{Q}_\tau^\star}\) can beat predicting the most likely class. The collapse is therefore a consequence of the assumption, not an unconditional fact. It captures the failure mode of relying on correlated but individually uninformative features once the truly predictive ones have been removed.

We next quantify the irreducible risk introduced by pruning informative features. Theorem~\ref{thm:psi-bound} establishes a gradient covariance lower bound on the error between the differentiable classifier \(f_\theta\) evaluated on the full feature vector and its best approximation using only the retained features.
\begin{quote}
\textbf{Assumption~2.} Let \(f_\theta:\mathbb{R}^d\to\mathbb{R}^C\) be a classifier with component logits \(\{f_\theta^{(c)}\}_{c=1}^C\). For class-conditional correlation-based pruning at a threshold \(\tau\in(0,1)\), write \(X=(Z,U)\), with \(Z\) the retained (kept) features and \(U\) the pruned features. Each logit is differentiable in \(U\) along the segment from \(\mu_y(Z):=\mathbb{E}[U\mid Z,Y=y]\) to \(U\), and \(\mathbb{E}[\|f_\theta(X)\|_2^2\mid Y=y]<\infty\), so that \(\mu_y(Z)\) and \(\Sigma_y^{-}(Z):=\operatorname{Cov}(U\mid Z,Y=y)\) exist.
\end{quote}

\begin{theorem}[Gradient based lower bound on functional irreducibility]
\label{thm:psi-bound}
Under Assumption~2, specify retained index sets \(\mathcal{Q}_\tau^{(y)}\subseteq[d]\) \((y\in[C])\), and construct \(\mathcal{Q}_\tau:=\bigcup_{y\in[C]}\mathcal{Q}_\tau^{(y)}\) and \(\mathcal{Q}_\tau^{-}:=[d]\setminus\mathcal{Q}_\tau\). For \(X\in\mathbb{R}^d\), write \(Z:=X_{\mathcal{Q}_\tau}\in\mathbb{R}^{|\mathcal{Q}_\tau|}\) and \(U:=X_{\mathcal{Q}_\tau^{-}}\in\mathbb{R}^{|\mathcal{Q}_\tau^{-}|}\).

Define the pruning-induced irreducibility $\Psi_\theta(\tau)$ as the supremum over classes of the minimal error when approximating $f_\theta(X)$ using only the retained features $Z$, with predictors $\hat{f}_\tau \in \mathcal{F}_\tau$ mapping $\mathbb{R}^{|\mathcal{Q}_\tau|} \to \mathbb{R}^C$:
\[
\Psi_\theta(\tau)\;:=\;\sup_{y\in[C]}\ \inf_{\hat{f}_\tau\in\mathcal{F}_\tau}\ \mathbb{E}\!\left[\big\|f_\theta(X)-\hat{f}_\tau(Z)\big\|_2^{2}\,\middle|\, Y=y\right].
\]

Then the following lower bound holds:
\begin{equation}
\label{eq:irreducibility}
\Psi_\theta(\tau)\;\ge\;\sup_{y\in[C]}\!\Bigl(\sqrt{L_y}-\sqrt{B_y}\Bigr)^{\!2},
\end{equation}
where, for each fixed class $y\in[C]$:
\begin{itemize}
\item $g_c(Z):=\nabla_U f_\theta^{(c)}\!\bigl(Z,\mu_y(Z)\bigr)$ is the $U$-gradient of the $c$-th logit evaluated at the anchor $\mu_y(Z)$ from Assumption~2.
\item $\Delta U:=U-\mu_y(Z)$ is the centered pruned block, and $u(t):=\mu_y(Z)+t\,\Delta U$ is the anchor-to-sample path.
\item $R_c(Z,U):=\int_0^1\!\bigl(\nabla_U f_\theta^{(c)}(Z,u(t))-g_c(Z)\bigr)^\top\!\Delta U\,dt$ is the pathwise remainder of the $c$-th logit along $u(t)$.
\item $L_y:=\mathbb{E}\!\Bigl[\sum_{c=1}^C g_c(Z)^\top\Sigma_y^-(Z)\,g_c(Z)\Bigm|Y=y\Bigr]$ is the expected gradient-weighted pruned-block variance (signal term).
\item $B_y:=\sum_{c=1}^C\mathbb{E}\!\Bigl[\bigl(R_c(Z,U)-\mathbb{E}[R_c(Z,U)\mid Z,Y=y]\bigr)^{\!2}\Bigm|Y=y\Bigr]$ is the total remainder variance across logits (curvature penalty).
\end{itemize}
\end{theorem}

\paragraph*{Proof} See Appendix~\ref{app:thm-psi}.

Inequality \eqref{eq:irreducibility} states that the pruning induced irreducibility $\Psi_\theta(\tau)$, the worst-case over classes of the smallest mean squared error achievable by any predictor that uses only the kept block $Z$, is at least the worst class gap $(\sqrt{L_y}-\sqrt{B_y})^{2}$. The signal term $L_y$ couples the sensitivity of the logits along the pruned directions $g_c(Z)$ with the residual covariance $\Sigma_y^-(Z)$ of the pruned block. The curvature term $B_y$ measures how far the classifier departs from linearity along the pruned directions, through the pathwise remainder $R_c(Z,U)$. When the pruned block is well explained by the kept block, $\Sigma_y^-(Z)$ is small and the bound is small. When the pruned block carries independent signal, the bound is large, which flags brittle pruning.

The curvature term admits a closed form under a smoothness assumption.
\begin{quote}
\textbf{Assumption~3 (H\"older smoothness).} For each class $y$, each logit $f_\theta^{(c)}$ is $\mathcal{C}^{1,\alpha}$ in $U$, that is its $U$ gradient is H\"older continuous of order $\alpha\in(0,1]$ with constant $H_{\alpha,y,c}\ge 0$, and $\mathbb{E}[\|U-\mu_y(Z)\|_2^{2+2\alpha}\mid Y=y]<\infty$.
\end{quote}

\begin{restatable}[H\"older curvature penalty]{corollary}{HolderGradientBound}
\label{cor:holder-gradient-bound}
Under the setting of Theorem~1 and Assumption~3, the pruning irreducibility satisfies
\begin{multline}
\label{eq:psi-holder-perc}
\Psi_\theta(\tau)\ \ge\ \sup_{y\in[C]}
\Bigg(
\sqrt{\underbrace{L_y}_{\text{signal term}}}
\\[3pt]
-\ \sqrt{\underbrace{\left(\sum_{c=1}^C \frac{H_{\alpha,y,c}^{2}}{(\alpha+1)^{2}}\right)
\mathbb{E}\!\Big[\|U-\mu_y(Z)\|_2^{\,2+2\alpha}\,\Big|\,Y=y\Big]}_{\text{penalty term}}}
\Bigg)^{\!2}.
\end{multline}
\end{restatable}

\paragraph*{Proof} See Appendix~\ref{app:cor-holder}.

The exponent $\alpha$ sets how fast the gradient may drift with distance from the anchor $\mu_y(Z)$. The constant $H_{\alpha,y,c}$ sets the amplitude of that drift and enters the penalty quadratically. The bound is trivial when $U$ is determined by $Z$, since then $\Sigma_y^-(Z)\equiv 0$ and $L_y\equiv 0$. It is tight in the affine case.

\begin{proposition}[Tightness in the affine case]
\label{prop:affine-segment}
Under the setting of Theorem~1, suppose that, for fixed \(z\), each logit \(f_\theta^{(c)}((z,u))\) is affine in \(u\) along the segment joining the anchor \(\mu_y(z)\) to \(u\). Then the pathwise remainder vanishes and hence \(B_y=0\). Consequently,
\begin{equation}
\label{eq:psi-affine-equality}
\Psi_\theta(\tau)\;=\;\sup_{y\in[C]} L_y.
\end{equation}
\end{proposition}

\paragraph*{Proof} See Appendix~\ref{app:prop-affine}.

When within-class correlation is high, the pruned block is well explained by the kept block, so $\Sigma_y^-(Z)$ remains small and the bound $(\sqrt{L_y}-\sqrt{B_y})^2$ is correspondingly small, confirming that the irreducible error from pruning stays low. When the pruned block carries independent class-conditional signal, $L_y$ grows and the bound rises as a quantitative warning that the pruning is brittle. Proposition~\ref{prop:affine-segment} shows the curvature penalty becomes trivial in the affine case. We verify the bound against empirical estimates in Subsection~\ref{subsec:validation}.

\section{Experimental Evaluation}
This section presents comparisons of downstream classification performance across CCMP and other supervised and class-conditional feature selectors, an ablation over cutoff-selection criteria, and bound verification, on high-dimensional spectroscopic and hyperspectral benchmarks.

\subsection{Datasets}
\label{subsec:datasets}
The four benchmarks were selected to span distinct measurement modalities, feature dimensions, and class counts, providing a test of CCMP that is not specific to any single sensor or regime. All four are publicly available. Strawberry~\cite{holland1998strawberry} is a Fourier transform infrared (FTIR) dataset of strawberry puree spectra from the UCR archive, evaluated on its canonical train/test partition. The remaining three use a stratified 75/25 train/test split. MRSA~\cite{yu2022mrsa} contains matrix-assisted laser desorption/ionization time-of-flight (MALDI-TOF) mass profiles of methicillin-resistant and susceptible clinical isolates. DRS Tissue~\cite{li2024tissue} covers eleven tissue classes collected via diffuse reflectance spectroscopy (DRS) at 355--1850\,nm. Botswana~\cite{ham2005randomforest} covers fourteen land-cover classes from 447--2355\,nm Hyperion spectra (145 bands retained after removing water-absorption regions). Table~\ref{tab:datasets} summarizes the datasets.

\begin{table}[t]
\centering
\caption{Benchmark datasets used for evaluation.}
\label{tab:datasets}
\setlength{\tabcolsep}{4pt}
\begin{tabular}{@{}llccc@{}}
\toprule
\textbf{Dataset} & \textbf{Modality} & \(\bm{N}\) & \(\bm{d}\) & \(\bm{C}\) \\
\midrule
Strawberry  & FTIR          & 983  & 235  & 2  \\
MRSA        & MALDI-TOF     & 2456 & 1000 & 2  \\
DRS Tissue  & DRS           & 7524 & 1531 & 11 \\
Botswana    & Hyperion      & 3248 & 145  & 14 \\
\bottomrule
\end{tabular}
\end{table}

\subsection{Evaluation Setup and Metrics}
\label{sec:classification}
Each feature subset (from CCMP and from each baseline) is input to logistic regression (LR), support vector machine (SVM), and random forest (RF) classifiers at fixed default settings via scikit-learn~\cite{pedregosa2011scikit}, and performance is reported as macro F1 averaged across classifiers. Baselines are evaluated at the same feature count \(k = |\widehat{\mathcal{Q}}_{\tau^\star}|\) as CCMP, so that any F1 difference comes from feature choice alone. Feature scaling (standardization to zero mean and unit variance), feature selection, and classifier training are performed on the training partition, and applied to the held-out test partition without modification, avoiding data leakage at any stage of the pipeline.

The evaluation metric is macro F1, which treats each class equally irrespective of sample count. For each class \(y\in[C]\), let \(\mathrm{TP}_y\), \(\mathrm{FP}_y\), and \(\mathrm{FN}_y\) denote the true positives, false positives, and false negatives. The per-class precision and recall are \(P_y = \mathrm{TP}_y/(\mathrm{TP}_y+\mathrm{FP}_y)\) and \(R_y = \mathrm{TP}_y/(\mathrm{TP}_y+\mathrm{FN}_y)\). We compute
\begin{equation}
\label{eq:macro-f1}
\mathrm{F1}
= \frac{1}{C}\sum_{y=1}^{C}\frac{2P_y R_y}{P_y+R_y+\varepsilon},
\end{equation}
where \(\varepsilon > 0\) is a small constant included to avoid division by zero.

\subsection{Feature Selector Comparison}
\label{subsec:comparison}

Table~\ref{tab:method_comparison} compares CCMP against class-conditional selectors (median absolute deviation / MAD, Laplacian score / LapScore, ridge reconstruction / RidgeRecon) and supervised selectors (Lasso, ElasticNet, \(\ell_1\)-SVM, Mutual Info.), where each baseline retains \(k\) features, the number CCMP selects on each dataset. It reports \(\overline{F1} := \tfrac{1}{3}(\mathrm{F1}_{\mathrm{LR}}+\mathrm{F1}_{\mathrm{SVM}}+\mathrm{F1}_{\mathrm{RF}})\), the mean macro-F1 across LR, SVM, and RF. The table also includes a full-feature baseline, a classifier trained on all \(d\) original features with no selection applied. \(\Delta_{\text{full}} := \overline{F1} - \overline{F1}_{\text{full}}\) measures the gap to the full-feature baseline, and \(\Delta_{\text{sup}} := \overline{F1} - \overline{F1}_{\text{sup}}^\star\) measures the gap to \(\overline{F1}_{\text{sup}}^\star\), the highest \(\overline{F1}\) among the supervised selectors. A box marks each of these two reference values in the table. A positive value means the row exceeds that reference. Bold marks CCMP, and underline marks the best value per column. Each column header gives the retained features \(k/d\): 55/235 (23\%) on Strawberry, 341/1000 (34\%) on MRSA, 42/1531 (3\%) on DRS Tissue, and 60/145 (41\%) on Botswana, ranging from 3\% retained on DRS Tissue to 41\% on Botswana. CCMP's \(|\Delta_{\text{full}}|\) reaches at most 1.7\% across the four datasets (DRS Tissue). Against class-conditional selectors, CCMP ranks first in every case, with the largest margin on DRS Tissue, where CCMP reaches \(86.9\%\) \(\overline{F1}\) while MAD, LapScore, and RidgeRecon collapse to \(53.6\%\), \(59.2\%\), and \(57.5\%\), respectively. Against supervised selectors, CCMP's \(|\Delta_{\text{sup}}|\) stays within \(1.2\%\) on Strawberry, MRSA, and Botswana. On DRS Tissue, CCMP surpasses every supervised method, reaching \(86.9\%\) \(\overline{F1}\) against the best supervised selector's \(84.2\%\) (ElasticNet).

\begin{table*}[t]
\centering
\caption{Macro-F1 (\%, one decimal) for supervised and class-conditional selectors on Strawberry, MRSA, DRS Tissue, and Botswana, each matched to CCMP's feature budget \(k=|\widehat{\mathcal{Q}}_{\tau^\star}|\). Columns report LR, SVM, RF, their mean \(\overline{F1}\) (Avg), and two gaps: \(\Delta_{\text{full}} := \overline{F1} - \overline{F1}_{\text{full}}\) to the full-feature baseline, and \(\Delta_{\text{sup}} := \overline{F1} - \overline{F1}_{\text{sup}}^\star\) to the best supervised selector, where a positive \(\Delta\) means the row beats that reference. \textbf{Bold} marks CCMP, \underline{underline} marks the best value per column, and \fbox{a box} marks the reference value used for each \(\Delta\).}
\label{tab:method_comparison}
\setlength{\tabcolsep}{1.2pt}
{\scriptsize
\begin{tabular}{@{}l cccccc @{\hspace{8.5pt}} cccccc @{\hspace{8.5pt}} cccccc @{\hspace{8.5pt}} cccccc@{}}
\toprule
& \multicolumn{6}{c}{\shortstack{\textbf{Strawberry FTIR (\%)} \\ \(d{=}235,\;k{=}55\)}}
& \multicolumn{6}{c}{\shortstack{\textbf{MRSA MALDI-TOF (\%)} \\ \(d{=}1000,\;k{=}341\)}}
& \multicolumn{6}{c}{\shortstack{\textbf{DRS Tissue (\%)} \\ \(d{=}1531,\;k{=}42\)}}
& \multicolumn{6}{c}{\shortstack{\textbf{Botswana Hyperion (\%)} \\ \(d{=}145,\;k{=}60\)}} \\
\cmidrule(lr){2-7}\cmidrule(lr){8-13}\cmidrule(lr){14-19}\cmidrule(lr){20-25}
\shortstack[l]{\textbf{Feature}\\\textbf{Selector}}
  & LR & SVM & RF & $\overline{F1}$ & $\Delta_{\text{full}}$ & $\Delta_{\text{sup}}$
  & LR & SVM & RF & $\overline{F1}$ & $\Delta_{\text{full}}$ & $\Delta_{\text{sup}}$
  & LR & SVM & RF & $\overline{F1}$ & $\Delta_{\text{full}}$ & $\Delta_{\text{sup}}$
  & LR & SVM & RF & $\overline{F1}$ & $\Delta_{\text{full}}$ & $\Delta_{\text{sup}}$ \\
\midrule
Full-feature
  & 95.1 & 93.7 & 96.5 & \fbox{95.1} & $0.0$ & $+0.4$
  & 75.3 & 80.8 & 78.6 & \fbox{78.3} & $0.0$ & $-0.8$
  & 95.2 & 82.1 & 88.6 & \fbox{88.6} & $0.0$ & $+4.4$
  & 96.9 & 91.9 & 93.5 & \fbox{94.1} & $0.0$ & $-0.1$ \\
\midrule
\multicolumn{25}{l}{\textit{Supervised selectors (matched at \(k\))}} \\[1pt]
Lasso
  & \underline{94.8} & \underline{93.7} & 95.6 & \fbox{\underline{94.7}} & \underline{$-0.4$} & \underline{$0.0$}
  & 74.6 & \underline{81.1} & 79.9 & 78.6 & $+0.3$ & $-0.5$
  & 85.1 & 78.2 & 86.0 & 83.1 & $-5.5$ & $-1.1$
  & \underline{96.5} & 92.5 & 93.8 & \fbox{\underline{94.3}} & \underline{$+0.1$} & \underline{$0.0$} \\
ElasticNet
  & 94.5 & \underline{93.7} & 95.6 & 94.6 & $-0.5$ & $-0.1$
  & 74.8 & 80.8 & \underline{81.6} & \fbox{\underline{79.1}} & \underline{$+0.8$} & \underline{$0.0$}
  & 85.5 & 80.9 & 86.3 & \fbox{84.2} & $-4.4$ & $0.0$
  & \underline{96.5} & \underline{92.6} & 93.7 & \fbox{\underline{94.3}} & \underline{$+0.1$} & \underline{$0.0$} \\
\(\ell_1\)-SVM
  & 94.2 & 93.1 & 95.6 & 94.3 & $-0.7$ & $-0.4$
  & 74.0 & 80.9 & 79.0 & 78.0 & $-0.3$ & $-1.1$
  & 83.3 & 77.3 & 84.5 & 81.7 & $-6.9$ & $-2.5$
  & 95.9 & 91.6 & \underline{94.0} & 93.8 & $-0.3$ & $-0.4$ \\
Mut.\ Info.
  & 91.6 & 92.3 & \underline{96.2} & 93.4 & $-1.7$ & $-1.3$
  & 76.4 & 79.2 & 80.0 & 78.5 & $+0.2$ & $-0.6$
  & 55.0 & 47.4 & 66.7 & 56.3 & $-32.3$ & $-27.9$
  & 93.6 & 85.3 & 89.2 & 89.4 & $-4.8$ & $-4.9$ \\
\midrule
\multicolumn{25}{l}{\textit{Class-conditional selectors (matched at \(k\))}} \\[1pt]
\textbf{CCMP (ours)}
  & \textbf{94.5} & \textbf{\underline{93.7}} & \textbf{95.4} & \textbf{94.5} & $\bm{-0.6}$ & $\bm{-0.2}$
  & \textbf{75.3} & \textbf{79.9} & \textbf{78.5} & \textbf{77.9} & $\bm{-0.4}$ & $\bm{-1.2}$
  & \textbf{\underline{90.6}} & \textbf{\underline{82.9}} & \textbf{\underline{87.3}} & \textbf{\underline{86.9}} & \underline{$\bm{-1.7}$} & \underline{$\bm{+2.7}$}
  & \textbf{95.8} & \textbf{92.0} & \textbf{93.7} & \textbf{93.8} & $\bm{-0.3}$ & $\bm{-0.4}$ \\
MAD
  & 92.2 & 87.1 & 91.2 & 90.2 & $-4.9$ & $-4.5$
  & 72.5 & 73.6 & 73.3 & 73.1 & $-5.2$ & $-6.0$
  & 55.9 & 45.6 & 59.3 & 53.6 & $-35.0$ & $-30.6$
  & 94.0 & 89.6 & 91.1 & 91.6 & $-2.6$ & $-2.7$ \\
LapScore
  & 92.5 & 83.4 & 88.9 & 88.3 & $-6.8$ & $-6.4$
  & \underline{77.6} & 77.3 & 76.4 & 77.1 & $-1.2$ & $-2.0$
  & 58.7 & 49.6 & 69.3 & 59.2 & $-29.4$ & $-25.0$
  & 95.1 & 90.2 & 92.8 & 92.7 & $-1.5$ & $-1.6$ \\
RidgeRecon
  & 92.1 & 90.0 & 92.9 & 91.7 & $-3.4$ & $-3.0$
  & 61.8 & 63.3 & 60.4 & 61.8 & $-16.5$ & $-17.3$
  & 52.4 & 49.3 & 70.7 & 57.5 & $-31.1$ & $-26.7$
  & 93.5 & 87.3 & 91.1 & 90.6 & $-3.5$ & $-3.6$ \\
\bottomrule
\end{tabular}
}
\end{table*}

\begin{figure}[!h]
\centering
\resizebox{\columnwidth}{!}{\includegraphics{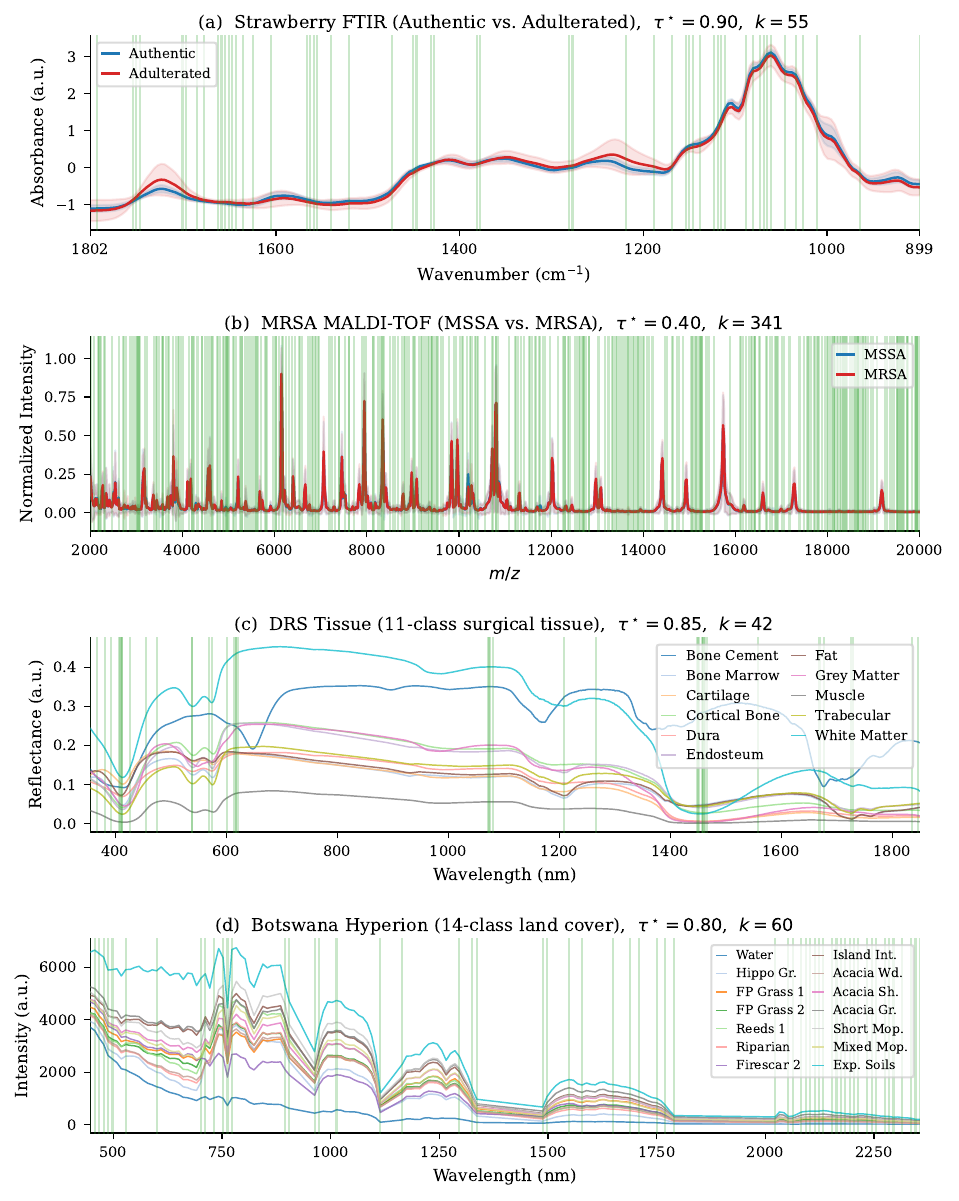}}
\caption{Each panel shows a dataset's per-class mean profile from the training split, with green shading marking the features CCMP retains. (a)~Strawberry FTIR (2 classes): $\tau^\star\!=\!0.90$, $k\!=\!55$ of 235 channels. (b)~MRSA MALDI-TOF (2 classes): $\tau^\star\!=\!0.40$, $k\!=\!341$ of 1000 bins. (c)~DRS Tissue (11 classes): $\tau^\star\!=\!0.85$, $k\!=\!42$ of 1531 channels. (d)~Botswana Hyperion (14 classes): $\tau^\star\!=\!0.80$, $k\!=\!60$ of 145 bands.}
\label{fig:interpretability}
\end{figure}

Fig.~\ref{fig:interpretability} highlights CCMP's retained feature positions with green shading over each dataset's per-class mean profile from the training split.

\subsection{Ablation over Cutoff-Selection Criteria}
\label{subsec:ablation}

\begin{figure}[!t]
\centering
\resizebox{\linewidth}{!}{%
\begin{tikzpicture}[
  b/.style={draw, rounded corners=2pt, minimum height=6mm, align=center, font=\footnotesize},
  abl/.style={draw, dashed, rounded corners=3pt, minimum height=10mm, align=center, font=\footnotesize, fill=orange!14},
  arr/.style={-{Stealth[scale=0.8]}, thick},
]
  \node[b, fill=gray!8,    minimum width=13mm]                     (data)  {$X_{\mathrm{tr}},\mathbf{y}$};
  \node[b, fill=blue!12,   minimum width=22mm, right=6mm of data]  (corr)  {per-class $|\boldsymbol{\mathcal{C}}^{(y)}|$};
  \node[b, fill=teal!10, minimum width=16mm, right=6mm of corr]          (tau)   {$\tau$ sweep};
  \node[b, fill=blue!12,   minimum width=20mm, right=6mm of tau]   (prune) {greedy prune\\per class};
  \node[b, fill=orange!14, minimum width=22mm, below=6mm of prune] (union) {union $\bigcup_y$};
  \node[abl, fill=teal!10, minimum width=34mm, left=6mm of union]  (crit)  {Criterion $\to$ $\tau^\star$\\{\scriptsize BIC\,$|$\,AIC\,$|$\,ELBOW}};
  \node[b, fill=green!12,  minimum width=14mm, left=6mm of crit]   (Q)     {$\widehat{\mathcal{Q}}_{\tau^\star}$};
  \node[b, fill=blue!8,    minimum width=14mm, left=6mm of Q]      (clf)   {Classifier};
  \draw[arr] (data)  -- (corr);
  \draw[arr] (corr)  -- (tau);
  \draw[arr] (tau)   -- (prune);
  \draw[arr] (prune) -- (union);
  \draw[arr] (union) -- (crit) node[midway, above, font=\scriptsize, text=gray!70!black] {$\widehat{\mathcal{Q}}_\tau$};
  \draw[arr] (crit)  -- (Q);
  \draw[arr] (Q)     -- (clf);
  \node[font=\scriptsize, text=gray!70!black, below=0.5mm of crit] {\textit{ablation point}};
\end{tikzpicture}%
}
\caption{Ablation pipeline. The dashed criterion block is the ablation point. Sweeping $\tau$ and pruning per class generates candidate feature sets, and the criterion selects $\tau^\star$ from BIC, AIC, or a geometric elbow rule. All other stages follow the proposed pipeline in Fig.~\ref{fig:overview_pipeline}.}
\label{fig:ablation_pipeline}
\end{figure}

This subsection compares three criteria, BIC~\cite{schwarz1978bic}, AIC~\cite{akaike1974new}, and the geometric elbow rule proposed in Subsection~\ref{sec:tau-selection}, for selecting the correlation cutoff \(\tau^\star\) from the swept \(\kappa(\tau)\) curve \eqref{eq:kappa-def}. Fig.~\ref{fig:ablation_pipeline} shows this ablation pipeline, with the criterion block as the single point of variation and every other stage fixed relative to Fig.~\ref{fig:overview_pipeline}. BIC selects
\begin{equation}
\label{eq:bic}
\tau^\star_{\mathrm{BIC}} = \arg\min_{\tau \in \mathcal{T}} \Big[-(d-k)\log\kappa(\tau) + k\log N\Big],
\end{equation}
and AIC replaces the \(k\log N\) penalty with a constant \(2k\),
\begin{equation}
\label{eq:aic}
\tau^\star_{\mathrm{AIC}} = \arg\min_{\tau \in \mathcal{T}} \Big[-(d-k)\log\kappa(\tau) + 2k\Big].
\end{equation}
Both share the fit term \(-(d-k)\log\kappa(\tau)\), which is large when many features are dropped while \(\kappa(\tau)\) remains far from 1, and vanishes as \(\kappa(\tau)\to1\) irrespective of how many features are dropped. The two penalties differ only in how they discourage keeping too many features. The BIC penalty, \(k\log N\), grows with the training sample size. The AIC penalty, \(2k\), stays constant. The elbow rule of Section~\ref{sec:tau-selection} uses neither term \eqref{eq:tau-star}.

\begin{figure}[t]
\centering
\includegraphics[width=\columnwidth]{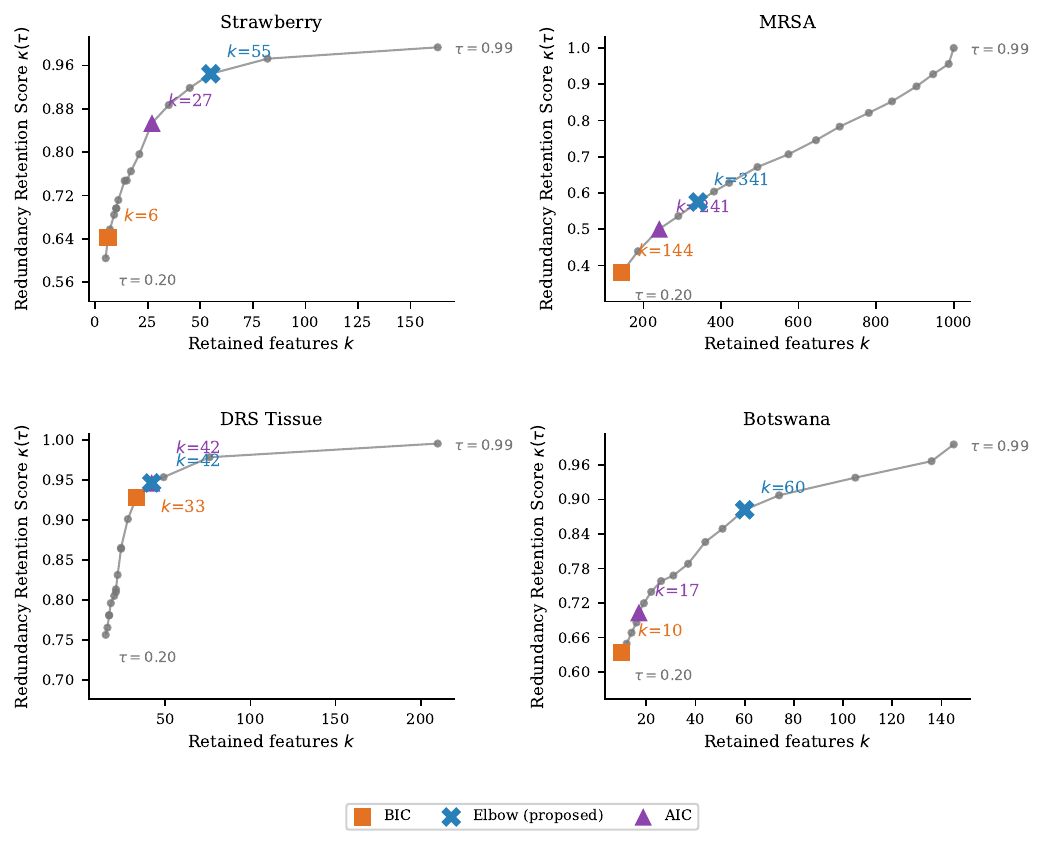}
\caption{Redundancy Retention Score $\kappa(\tau)$ against the retained-feature count $k$ for each dataset, swept over the candidate grid $\mathcal{T} = \{0.20,0.25,\dots,0.95,0.99\}$. The gray labels mark the two endpoints of this sweep, $\tau=0.20$ and $\tau=0.99$. Markers show where BIC, AIC, and the geometric elbow rule each select $\tau^\star$ along this curve.}
\label{fig:elbow_curve}
\end{figure}

Fig.~\ref{fig:elbow_curve} plots \(\kappa(\tau)\) against \(k\) for each dataset. Lower \(\tau\) values in \(\mathcal{T}\) identify more pairs as correlated \eqref{eq:pair_set}, so pruning discards more features and \(k\) shrinks, while higher \(\tau\) values identify fewer pairs and let more features remain. The three markers on each curve show where BIC, AIC, and the elbow rule each stop this sweep. BIC's penalty grows with \(N\) and stops the sweep earliest. MRSA's \(\kappa(\tau)\) curve is the exception. The other three datasets show a sharp rise in \(\kappa(\tau)\) before leveling off, but MRSA's rises gradually across the entire sweep. No criterion finds a clear cutoff, though the elbow rule still reaches the highest value of the three.

\begin{figure}[t]
\centering
\includegraphics[width=\columnwidth]{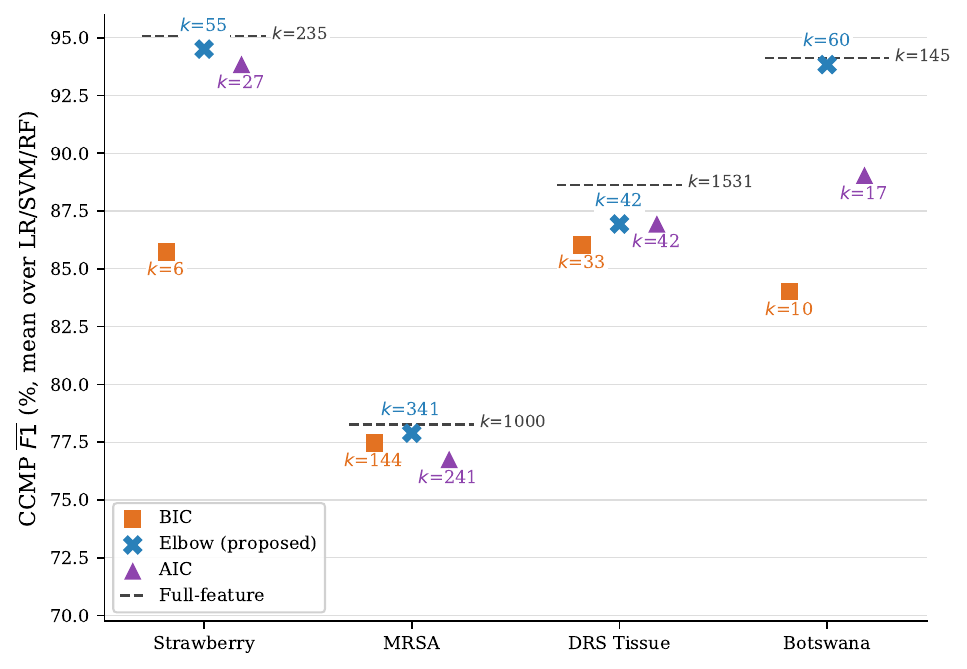}
\caption{Criterion ablation. $\overline{F1}$ is the mean over LR, SVM, and RF. For each dataset, BIC (orange square), the proposed geometric elbow (blue cross), and AIC (purple triangle) each select a $\tau^\star$ and retained-feature budget $k$, annotated above or below each marker. The dashed line marks the full-feature baseline, labeled with the total (unpruned) feature count $k{=}d$. The elbow criterion's marker remains closest to this baseline across all four datasets.}
\label{fig:ablation}
\end{figure}

Fig.~\ref{fig:ablation} compares each criterion's \(\overline{F1}\) at its own \(\tau^\star\) against the full-feature baseline. On Strawberry, BIC keeps only 6 of 235 features, dropping \(\overline{F1}\) to \(85.7\%\), versus the elbow rule's \(94.5\%\) at \(k=55\). The elbow rule is closest to full-feature performance on three datasets, matching AIC on DRS Tissue at \(k=42\).

\subsection{Empirical Validation of the Irreducibility Bound}
\label{subsec:validation}

This subsection tests Theorem~\ref{thm:psi-bound} against empirical data. At \(\tau^\star\), CCMP fixes which features are kept and which are pruned. \(\mathrm{LHS}_{\sup}=\max_y \mathrm{LHS}_y\) is the minimal error of predicting any differentiable classifier's output using only the retained features, estimated by Monte Carlo, the error pruning cannot avoid. \(\mathrm{RHS}_{\sup}=\max_y(\sqrt{\widehat L_y}-\sqrt{\widehat B_y})^2\) is Theorem~\ref{thm:psi-bound}'s lower bound on that error, the pruned features' gradient-weighted variance minus a curvature penalty from the classifier's H\"older smoothness, evaluated across exponents \(\alpha\in\{1/4,1/2,3/4,1\}\). \(\mathrm{LHS}_{\sup}\) meets or exceeds \(\mathrm{RHS}_{\sup}\) in all sixteen configurations tested, four datasets and four H\"older exponents, confirming the assumption behind the bound for high-dimensional, correlated measurements. This bound, however, is not specific to \(\tau^\star\). It holds for any split \((Z,U)\) at any \(\tau\) satisfying Assumption~2. Table~\ref{tab:validation} reports the margin between \(\mathrm{LHS}_{\sup}\) and \(\mathrm{RHS}_{\sup}\) for every configuration, positive whenever the bound holds, and Appendix~\ref{app:bound_evaluation} outlines the estimation procedure.

\begin{table}[!t]
\centering
\caption{Empirical validation of the irreducibility bound (Theorem~\ref{thm:psi-bound}) using the closed-form H\"older penalty from Corollary~\ref{cor:holder-gradient-bound}. For each dataset and H\"older exponent \(\alpha\), Monte Carlo \(\mathrm{LHS}_{\sup}\) is compared with analytical \(\mathrm{RHS}_{\sup}\). A nonnegative margin confirms the bound holds.}
\label{tab:validation}
\setlength{\tabcolsep}{4pt}
\begin{tabular}{@{}llccc@{}}
\toprule
\textbf{Dataset} & \(\bm{\alpha}\) & \(\mathbf{LHS_{\sup}}\) & \(\mathbf{RHS_{\sup}}\) & \textbf{Margin} \\
\midrule
Strawberry  & 1/4 & \(1.20\e{-1}\) & \(1.03\e{-1}\) & \(+1.73\e{-2}\) \\
Strawberry  & 1/2 & \(1.20\e{-1}\) & \(1.05\e{-1}\) & \(+1.53\e{-2}\) \\
Strawberry  & 3/4 & \(1.20\e{-1}\) & \(1.06\e{-1}\) & \(+1.40\e{-2}\) \\
Strawberry  & 1   & \(1.20\e{-1}\) & \(1.07\e{-1}\) & \(+1.33\e{-2}\) \\
\midrule
MRSA        & 1/4 & \(2.87\e{0}\)  & \(1.03\e{-1}\) & \(+2.77\e{0}\)  \\
MRSA        & 1/2 & \(2.87\e{0}\)  & \(2.85\e{-1}\) & \(+2.59\e{0}\)  \\
MRSA        & 3/4 & \(2.88\e{0}\)  & \(3.12\e{-1}\) & \(+2.56\e{0}\)  \\
MRSA        & 1   & \(2.88\e{0}\)  & \(1.31\e{-1}\) & \(+2.75\e{0}\)  \\
\midrule
DRS Tissue  & 1/4 & \(1.14\e{-1}\) & \(8.91\e{-2}\) & \(+2.50\e{-2}\) \\
DRS Tissue  & 1/2 & \(1.14\e{-1}\) & \(9.54\e{-2}\) & \(+1.89\e{-2}\) \\
DRS Tissue  & 3/4 & \(1.14\e{-1}\) & \(9.85\e{-2}\) & \(+1.58\e{-2}\) \\
DRS Tissue  & 1   & \(1.14\e{-1}\) & \(9.87\e{-2}\) & \(+1.56\e{-2}\) \\
\midrule
Botswana    & 1/4 & \(2.05\e{-1}\) & \(1.78\e{-1}\) & \(+2.66\e{-2}\) \\
Botswana    & 1/2 & \(2.05\e{-1}\) & \(1.82\e{-1}\) & \(+2.26\e{-2}\) \\
Botswana    & 3/4 & \(2.05\e{-1}\) & \(1.85\e{-1}\) & \(+1.93\e{-2}\) \\
Botswana    & 1   & \(2.04\e{-1}\) & \(1.87\e{-1}\) & \(+1.72\e{-2}\) \\
\bottomrule
\end{tabular}
\end{table}

For MRSA, \(\mathrm{LHS}_{\sup}\) exceeds the other three datasets by two orders of magnitude, while \(\mathrm{RHS}_{\sup}\) stays comparable across all four. MRSA's discrete m/z peaks correlate weakly with the kept features, leaving the pruned block poorly explained by the kept block. This inflates \(\Sigma_y^-(Z)\) and the signal term \(L_y\) (Subsection~\ref{subsec:theoretical-guarantees}), which accounts for the large \(\mathrm{LHS}_{\sup}\). Table~\ref{tab:method_comparison} also confirms CCMP's MRSA \(\overline{F1}\) is the lowest of the four datasets.

\section{Discussion}
\label{subsec:discussion}

CCMP prunes one feature from each strongly correlated pair, per class, and substantially reduces the feature count. It outperforms every class-conditional baseline across all four datasets, and its $\overline{F1}$ falls at most 1.7\% below the full-feature model (Section~\ref{subsec:ablation}). On DRS Tissue dataset, CCMP achieves $\overline{F1}=86.9\%$ while retaining only 42 of 1531 features (3\%), above the best supervised selector's $84.2\%$ (ElasticNet, Table~\ref{tab:method_comparison}). The theoretical bounds of Theorem~\ref{thm:psi-bound} and Corollary~\ref{cor:holder-gradient-bound} formally characterize the error induced by pruning, and all sixteen empirical evaluations of these bounds pass (Table~\ref{tab:validation}).

CCMP is competitive with supervised selection where within-class multicollinearity is the primary correlation structure, since it targets same-class redundancy, not between-class discriminability. However, MRSA is the exception. Its MALDI-TOF mass spectrum exhibits sharp spikes at discrete m/z positions, keeping same-class correlation low ($\tau^\star = 0.40$). At this low cutoff, correlation above $0.40$ is weak evidence of redundancy. CCMP still prunes on it, discarding some class-discriminative signal and letting supervised selectors edge ahead by a small margin in $\overline{F1}$ (ElasticNet $79.1\%$, CCMP $77.9\%$, Table~\ref{tab:method_comparison}).

\section{Conclusion}

This paper investigated how well pruning correlated features per class competes with supervised selectors, class-conditional selectors, and the unpruned full-feature baseline. CCMP beats every class-conditional baseline across four measurement modalities, and on the tissue dataset, retaining only $3\%$ of the original features, it exceeds supervised selection outright. Elsewhere, CCMP's gap to supervised selection stays within $1.2\%$, consistent with supervised methods drawing on class-discriminative information that CCMP does not use. The current formulation targets only linear dependencies, captured by Pearson correlation. Generalizing the pruning criterion to nonlinear and higher-order dependencies would let CCMP prune additional redundant features.

\section*{Acknowledgments}
The authors thank Samin Riasat (Northeastern University, Boston, MA) for thoughtful feedback and discussions on the mathematical analysis, particularly the theoretical guarantees in Subsection~\ref{subsec:theoretical-guarantees}. The authors used Claude Sonnet 5 (Anthropic, 2026) to edit the manuscript for clarity and readability, and to assist with the implementation of the experiments in Subsection~\ref{subsec:validation}. All AI assisted content was reviewed, fact checked, and revised by the authors, who take full responsibility for the content and its integrity.

\appendices

\section{Proof of Proposition~1}
\label{app:prop-bayes}

\begin{proof}
The conditional distribution of \( Y \) depends only on the subvector \( X_{\mathcal{Q}_\tau^\star} \); that is,
\begin{align}
    \mathbb{P}(Y \mid X) = \mathbb{P}(Y \mid X_{\mathcal{Q}_\tau^\star}) \; \Longleftrightarrow \; Y \perp X_{[d] \setminus \mathcal{Q}_\tau^\star} \mid X_{\mathcal{Q}_\tau^\star}.
\end{align}

Removing \( \mathcal{Q}_\tau^\star \) leaves only the uninformative subset \( \overline{\mathcal{Q}_\tau^\star} = [d] \setminus \mathcal{Q}_\tau^\star \), which conveys no information about \( Y \), so
\begin{equation}
    \mathbb{P}(Y \mid X_{\overline{\mathcal{Q}_\tau^\star}}) = \mathbb{P}(Y),
\end{equation}
i.e., \( Y \) is marginally independent of the retained features.

Consequently, no classifier relying solely on \( X_{\overline{\mathcal{Q}_\tau^\star}} \) can outperform constant prediction. The Bayes-optimal classifier is the constant rule
\begin{equation}
    \phi^\star(x) \equiv \arg\max_{c \in [C]} \mathbb{P}(Y = c),
\end{equation}
which achieves the smallest possible misclassification probability (the \emph{Bayes risk}) among all classifiers using only the retained features \(X_{\overline{\mathcal{Q}_\tau^\star}}\).
The corresponding Bayes risk is
\begin{equation}
    \inf_{\phi} \mathbb{P}(\phi(X_{\overline{\mathcal{Q}_\tau^\star}}) \ne Y)
= 1 - \max_{c \in [C]} \mathbb{P}(Y = c),
\end{equation}
as given by the classical Bayesian decision theory~\cite{VanTrees2001, lehmann2022testing}.
\end{proof}

\section{Proof of Theorem~1}
\label{app:thm-psi}

\begin{proof}
Under the setting of Theorem~\ref{thm:psi-bound}, write \(p=|\mathcal{Q}_\tau|\) and \(q=|\mathcal{Q}_\tau^-|\). With \(X\) partitioned as \((Z,U)\), we write
\begin{equation}
\label{eq:logit-block}
f_\theta^{(c)}(X)=f_\theta^{(c)}(Z,U),\qquad c\in\{1,\dots,C\}.
\end{equation}

For each \(y\in[C]\), consider mappings \(\hat{f}_\tau:\mathbb{R}^p\to\mathbb{R}^C\) with \(\mathbb{E}[\|\hat{f}_\tau(Z)\|_2^2\,|\,Y=y]<\infty\). By the \(L^2\) projection theorem, the optimal \(Z\)-only predictor is \(\hat{f}_{\tau,y}^\star(Z)=\mathbb{E}[f_\theta(X)\mid Z, Y=y]\in\mathbb{R}^C\), and the minimal value equals the total residual conditional variance across logits:
\begin{align}
&\inf_{\hat{f}_\tau\in\mathcal{F}_\tau}
\ \mathbb{E}\!\left[\left\|f_\theta(X)-\hat{f}_\tau(Z)\right\|_2^2 \,\middle|\, Y=y\right] \notag\\
&= \mathbb{E}\!\left[\left\|f_\theta(X)-\hat{f}_{\tau,y}^\star(Z)\right\|_2^2 \,\middle|\, Y=y\right] \notag\\
&= \mathbb{E}\!\left[\sum_{c=1}^C \Big(f_\theta^{(c)}(X)-\mathbb{E}\!\left[f_\theta^{(c)}(X)\mid Z, Y=y\right]\Big)^{\!2} \,\middle|\, Y=y\right] \notag\\
&= \sum_{c=1}^C \mathbb{E}\!\left[\operatorname{Var}\!\left(f_\theta^{(c)}(X)\,\middle|\, Z, Y=y\right)\right].
\label{eq:step1-varsum}
\end{align}

On the event \(Y=y\), introduce the pruned-block anchor as the conditional mean
\begin{equation}
\label{eq:mu-anchor}
\mu_y(Z):=\mathbb{E}\!\left[\,U \,\middle|\, Z,\, Y=y\right]\in\mathbb{R}^{q}.
\end{equation}

Now fix \(z\in\mathbb{R}^{p}\) (a realization of \(Z\)) and replace \(U\) by its anchor \(\mu_y(z)\), forming the anchored input \((z,\mu_y(z))\in\mathbb{R}^d\).
Traverse the straight line in the pruned subspace from \(\mu_y(z)\) to the observed \(U\) via
\begin{equation}
\label{eq:anchor-path}
u(t)\;:=\;\mu_y(z)+t\,\Delta U,\qquad \Delta U:=U-\mu_y(z),\quad t\in[0,1].
\end{equation}

Define the scalar path (per logit)
\begin{equation}
h_c(t)\;:=\;f_\theta^{(c)}\!\big(z,\,u(t)\big)\;=\;f_\theta^{(c)}\!\big(z,\,\mu_y(z)+t\,\Delta U\big),
\end{equation}
and assume \(f_\theta^{(c)}\) is differentiable in \(U\) along this segment (so \(h_c\) is absolutely continuous). Then
\begin{equation}
h_c(0)=f_\theta^{(c)}\!\big(z,\mu_y(z)\big),\qquad
h_c(1)=f_\theta^{(c)}(z,U).
\end{equation}

We consider the block gradients with respect to the kept and pruned indices:
\begin{align}
\label{eq:block-grads}
\nabla_Z f_\theta^{(c)}(z,u(t))
&:= \begin{bmatrix}
\frac{\partial f_\theta^{(c)}}{\partial z_1}(z,u(t))\\[-2pt]
\vdots\\[-2pt]
\frac{\partial f_\theta^{(c)}}{\partial z_p}(z,u(t))
\end{bmatrix}\in\mathbb{R}^{p},
\\[6pt]
\nabla_U f_\theta^{(c)}(z,u(t))
&:= \begin{bmatrix}
\frac{\partial f_\theta^{(c)}}{\partial u_1}(z,u(t))\\[-2pt]
\vdots\\[-2pt]
\frac{\partial f_\theta^{(c)}}{\partial u_q}(z,u(t))
\end{bmatrix}\in\mathbb{R}^{q}.
\end{align}

By the Fundamental Theorem of Calculus and the chain rule,
\begin{multline}
\label{eq:FTC-chain}
f_\theta^{(c)}(z,U) - f_\theta^{(c)}\!\big(z,\mu_y(z)\big)
= \int_0^1 \frac{d}{dt}\,h_c(t)\,dt
\\[2pt]
= \int_0^1 \nabla_Z f_\theta^{(c)}\!\big(z,u(t)\big)^\top \underbrace{\tfrac{d}{dt}z}_{=\,0}\,dt
\\[2pt]
\;+\; \int_0^1 \nabla_U f_\theta^{(c)}\!\big(z,u(t)\big)^\top \underbrace{\tfrac{d}{dt}u(t)}_{=\,\Delta U}\,dt \\[2pt]
= \int_0^1 \nabla_U f_\theta^{(c)}\!\big(z,u(t)\big)^\top \Delta U\,dt
\\[2pt]
= \int_0^1 \Big\langle \nabla_U f_\theta^{(c)}\!\big(z,\mu_y(z)+t\,\Delta U\big),\, \Delta U \Big\rangle dt \\[2pt]
= \underbrace{\Big\langle \nabla_U f_\theta^{(c)}\!\big(z,\mu_y(z)\big),\,\Delta U \Big\rangle}_{\text{anchor (linear) term}}
\\[2pt]
\;+\;
\underbrace{\int_0^1 \Big\langle \nabla_U f_\theta^{(c)}\!\big(z,u(t)\big)-\nabla_U f_\theta^{(c)}\!\big(z,\mu_y(z)\big),\, \Delta U \Big\rangle dt}_{R_c(z,U)}.
\end{multline}

Define the anchor gradient \(g_c(z):=\nabla_{U} f_\theta^{(c)}(z,\mu_y(z))\in\mathbb{R}^{q}\) is the direction of steepest increase in the \(U\)-subspace at \((z,\mu_y(z))\), and
\(g_c(z)^\top \Delta U\) gives the directional (first-order) change of the \(c\)-th logit toward \(U\). The remainder \(R_c(z,U)\in\mathbb{R}\), which measures how the local \(U\)-gradients along the path \(u(t)\) deviate from \(g_c(z)\), is integrated and projected onto \(\Delta U\), yielding
\begin{equation}
\label{eq:lin-plus-rem-z}
\begin{aligned}
f_\theta^{(c)}(z,U)
&= f_\theta^{(c)}\!\big(z,\mu_y(z)\big)
\;+ \underbrace{g_c(z)^\top \Delta U}_{\text{linear at anchor}}
+ \underbrace{R_c(z,U)}_{\text{integral remainder}} ,
\end{aligned}
\end{equation}
where,
\begin{equation}
\label{eq:R-def-z}
R_c(z,U)
= \int_0^1\!
\Big\langle
  \nabla_{U} f_\theta^{(c)}\!\big(
      z,
      \mu_y(z)+t\,\!\Delta U
  \big) - g_c(z),\ 
  \!\Delta U
\Big\rangle dt.
\end{equation}

Applying \(\mathbb{E}[\cdot\,|\,Z=z,\,Y=y]\) to \eqref{eq:lin-plus-rem-z} yields
\begin{multline}
\label{eq:cond-mean-zy}
\mathbb{E}\!\big[ f_\theta^{(c)}(z,U) \,\big|\, Z=z,\, Y=y \big]
\\
= f_\theta^{(c)}\!\big( z, \mu_y(z) \big)
+  g_c(z)^\top\, \underbrace{\mathbb{E}\!\big[ U-\mu_y(z) \,\big|\, Z=z,\, Y=y \big] }_{=\,0}
\\
{}+ \mathbb{E}\!\big[ R_c(z,U) \,\big|\, Z=z,\, Y=y \big]
\\
= f_\theta^{(c)}\!\big( z, \mu_y(z) \big)
+ \mathbb{E}\!\big[ R_c(z,U) \,\big|\, Z=z,\, Y=y \big]. 
\end{multline}

Subtracting \eqref{eq:cond-mean-zy} from \eqref{eq:lin-plus-rem-z} gives
\begin{multline}
\label{eq:centered-decomp-zy-z}
f_\theta^{(c)}(z,U)
- \mathbb{E}\!\big[f_\theta^{(c)}(z,U)\,\big|\,Z=z,\,Y=y\big]
\\
= g_c(z)^\top\!\Delta U
+ \Big(R_c(z,U)
- \mathbb{E}\!\big[R_c(z,U)\,\big|\,Z=z,\,Y=y\big]\Big).
\end{multline}

Using the triangle inequality (Minkowski) in the conditional \(L^2\) norm with
\(a:=f_\theta^{(c)}(z,U)-\mathbb{E}[f_\theta^{(c)}(z,U)\mid Z=z,\,Y=y]\) and
\(b:=R_c(z,U)-\mathbb{E}[R_c(z,U)\mid Z=z,\,Y=y]\), we have
\(\|a-b\|_{2\mid Z=z,\,Y=y}\le \|a\|_{2\mid Z=z,\,Y=y}+\|b\|_{2\mid Z=z,\,Y=y}\),
hence \(\|a\|_{2\mid Z=z,\,Y=y}\ge \|a-b\|_{2\mid Z=z,\,Y=y}-\|b\|_{2\mid Z=z,\,Y=y}\).
Substituting \(a-b=g_c(z)^\top\Delta U\) from \eqref{eq:centered-decomp-zy-z} gives
\begin{multline}
\label{eq:minkowski-derive-zy-z}
\big\|\,f_\theta^{(c)}(z,U)-\mathbb{E}[f_\theta^{(c)}(z,U)\mid Z=z,\,Y=y]\,\big\|_{2\mid Z=z,\,Y=y}
\\
\ge\ \big\|\,g_c(z)^\top \Delta U\,\big\|_{2\mid Z=z,\,Y=y} 
\\
\ -\ \big\|\,R_c(z,U)-\mathbb{E}[R_c(z,U)\mid Z=z,\,Y=y]\,\big\|_{2\mid Z=z,\,Y=y}\, 
\end{multline}

Here, $g_c(Z)^\top \Delta U$ has mean zero (since \(\mathbb{E}[\Delta U\mid Z,Y=y]=\mathbf{0}\)), and
\(
\Sigma_y^-(Z):=\mathbb{E}[\Delta U\,\Delta U^\top\mid Z,Y=y]
\). Its conditional second moment reduces to:
\begin{align}
&\big\|\,g_c(z)^\top \Delta U\,\big\|_{2\mid Z=z,\,Y=y}
\notag \\
&= \sqrt{\,\mathbb{E}\!\big[(g_c(z)^\top \Delta U)^2 \,\big|\, Z=z,\,Y=y\big]\,}
\notag \\
&= \sqrt{\,\mathbb{E}\!\big[\Delta U^\top\, g_c(z)g_c(z)^\top\, \Delta U \,\big|\, Z=z,\,Y=y\big]\,}
\notag \\
&= \sqrt{\,\operatorname{tr}\!\Big(g_c(z)g_c(z)^\top\, \mathbb{E}\!\big[\Delta U\,\Delta U^\top \,\big|\, Z=z,\,Y=y\big]\Big)\,}
\notag \\
&= \sqrt{\,g_c(z)^\top\, \Sigma_y^-(z)\, g_c(z)\,}\,.
\end{align}

Note that \(\|a\|_{2\mid Z=z,Y=y}^2=\mathbb{E}[a^2\mid Z=z,Y=y]=\operatorname{Var}(f_\theta^{(c)}(z,U)\mid Z=z,Y=y)\), since \(a\) is the centered logit under the same conditioning. Hence squaring \eqref{eq:minkowski-derive-zy-z} yields 
\begin{multline}
\label{eq:var-lb-star}
\operatorname{Var}\!\big(f_\theta^{(c)}(z,U)\mid Z=z, Y=y\big)
 \ge\
\Big(\sqrt{g_c(z)^\top\Sigma_y^-(z)g_c(z)}
\\
-\sqrt{\mathbb{E}\big[(R_c
-\mathbb{E}[R_c\mid Z=z,Y=y])^2\mid Z=z,Y=y\big]} \ \Big)^{2},
\end{multline}
where \(R_c := R_c(z,U)\).

Applying $\mathbb{E}[\cdot\mid Y=y]$ over the distribution of $Z$ given $Y=y$ and summing across all logits $c$ in $\eqref{eq:var-lb-star}$ yields
\begin{align}
\label{eq:sum-per-logit}
\sum_{c=1}^C\mathbb{E}\!\left[\mathrm{Var}\!\big(f_\theta^{(c)}(Z,U)\,\big|\,Z,Y=y\big)\,\Big|\,Y=y\right]
\notag \\
\ge\
\mathbb{E}\!\left[\sum_{c=1}^C
\!\Big(\!\sqrt{a_c(Z)}\!-\!\sqrt{b_c(Z)}\!\Big)^{\!2}\,\Big|\,Y=y\right],
\end{align}
where, for brevity,
\begin{equation}
\begin{aligned}
\label{eq:ab-def}
a_c(Z)&:= g_c(Z)^\top \Sigma_y^-(Z)\, g_c(Z)\ \ge 0,
\\
b_c(Z)&:= \mathbb{E}\!\big[\,(R_c-\mathbb{E}[R_c\mid Z,Y=y])^{2}\,\big|\, Z,Y=y\big]\ \ge 0.
\end{aligned}
\end{equation}

Applying Cauchy–Schwarz twice: (i) pointwise in \(Z\) on \(\mathbb{R}^C\),
\(\big(\sum_{c=1}^C \sqrt{a_c(Z)b_c(Z)}\big)^{2} \le (\sum_{c=1}^C a_c(Z))(\sum_{c=1}^C b_c(Z))\),
and (ii) in \(L^2\) on \(\sqrt{\sum_{c=1}^C a_c(Z)}\) and \(\sqrt{\sum_{c=1}^C b_c(Z)}\) under \(Y=y\), i.e.,
\(\mathbb{E}\!\big[\sqrt{(\sum_{c=1}^C a_c(Z))(\sum_{c=1}^C b_c(Z))}\,\big|\,Y=y\big]
\le \sqrt{\mathbb{E}[\sum_{c=1}^C a_c(Z)\mid Y=y]\,\mathbb{E}[\sum_{c=1}^C b_c(Z)\mid Y=y]}\), we obtain:
{\allowdisplaybreaks
\begin{multline}
\label{eq:one-line-chain}
\sum_{c=1}^C \mathbb{E}\!\big[\mathrm{Var}\big(f_\theta^{(c)}(Z,U)\mid Z,\,Y=y\big)\,\big|\,Y=y\big]
\\
\ge\;
\mathbb{E}\!\left[\sum_{c=1}^C \big(\sqrt{a_c(Z)}-\sqrt{b_c(Z)}\,\big)^{2}\,\Big|\,Y=y\right]
\\
=\;
\mathbb{E}\!\left[\sum_{c=1}^C a_c(Z)+\sum_{c=1}^C b_c(Z)\,\Big|\,Y=y\right]
\\
\;-\;2\,\mathbb{E}\!\left[\sum_{c=1}^C \sqrt{a_c(Z)b_c(Z)}\,\Big|\,Y=y\right]
\\
\ge\;
\mathbb{E}\!\left[\sum_{c=1}^C a_c(Z)+\sum_{c=1}^C b_c(Z)\,\Big|\,Y=y\right]
\\
\;-\;2\,\mathbb{E}\!\left[\sqrt{\left(\sum_{c=1}^C a_c(Z)\right)\left(\sum_{c=1}^C b_c(Z)\right)}\,\Big|\,Y=y\right]
\\
\ge\;
\sum_{c=1}^C \mathbb{E}[a_c(Z)\mid Y=y]
+\sum_{c=1}^C \mathbb{E}[b_c(Z)\mid Y=y]
\\
\;-\;2\,\sqrt{\left(\sum_{c=1}^C \mathbb{E}[a_c(Z)\mid Y=y]\right)\left(\sum_{c=1}^C \mathbb{E}[b_c(Z)\mid Y=y]\right)}
\\
=\;
\left(\sqrt{\sum_{c=1}^C \mathbb{E}[a_c(Z)\mid Y=y]}-\sqrt{\sum_{c=1}^C \mathbb{E}[b_c(Z)\mid Y=y]}\right)^{2}.
\end{multline}
}

Define the class-wise aggregates
\begin{flalign}
L_y :&= \sum_{c=1}^C \mathbb{E}[a_c(Z)\mid Y=y] & \notag \\
    &= \mathbb{E}\!\left[\sum_{c=1}^C g_c(Z)^\top \Sigma_y^-(Z)\, g_c(Z)\,\Big|\,Y=y\right],&
\label{eq:L-def} \\
B_y &:=\!\sum_{c=1}^C \mathbb{E}[b_c(Z)\mid Y=y] &
     \notag \\
    &=\!\sum_{c=1}^C\!\mathbb{E}\!\left[\mathbb{E}\!\left[(R_c-\!\mathbb{E}[R_c\!\mid\! Z,Y=y])^{2}\,\Big|\, Z,Y=y\right]\Big|Y=y\right] &
    \notag \\
    &=\!\sum_{c=1}^C\!\mathbb{E}\!\left[(R_c - \!\mathbb{E}[R_c\!\mid\! Z,Y=y])^{2}\Big|Y=y\right]. & \label{eq:B-def}
\end{flalign}

Thus,
\begin{align}
\label{eq:class-bound}
\sum_{c=1}^C \mathbb{E}\!\big[\mathrm{Var}\big(f_\theta^{(c)}(Z,U)\mid Z,\,Y=y\big)\,\Big|\,Y=y\big]
\notag \\
\;\ge\;
\big(\sqrt{L_y}-\sqrt{B_y}\,\big)^{2}.
\end{align}

The functional irreducibility $\Psi_\theta(\tau)$ is the supremum over classes of the per-class approximation error from \eqref{eq:step1-varsum}:
\begin{align}
\Psi_\theta(\tau)
&:= \sup_{y\in[C]} \sum_{c=1}^C \mathbb{E}\!\left[\mathrm{Var}\!\big(f_\theta^{(c)}(Z,U)\mid Z,\,Y=y\big)\,\Big|\,Y=y\right] \notag\\
&\ge \sup_{y\in[C]}\!\left(\sqrt{L_y}-\sqrt{B_y}\right)^{2}.
\label{eq:psi-bound}
\end{align}
This is the claimed inequality \eqref{eq:irreducibility}.

\end{proof}

\section{Proof of Corollary~1}
\label{app:cor-holder}

\begin{proof}
Consider the straight path \(u(t)=\mu_y(z)+t\,\Delta U\) for \(t\in[0,1]\), where \(u(0)=\mu_y(z)\) is the anchor and \(g_c(z)=\nabla_U f_\theta^{(c)}(z,u(0))\). Decompose \(\Delta U = r\,\widehat{\Delta U}\) with \(r:=\|\Delta U\|_2\) and \(\widehat{\Delta U}:=\Delta U/\|\Delta U\|_2\) (when \(r>0\)), so \(\widehat{\Delta U}\) is the unit direction. Define the gradient difference
\begin{equation}
\label{eq:Delta-nabla-def}
\Delta\nabla_c(t)\;:=\;\nabla_U f_\theta^{(c)}(z,u(t))-\nabla_U f_\theta^{(c)}(z,u(0))\in\mathbb{R}^q.
\end{equation}

Assume the \(U\)-gradient has a class-conditional modulus of continuity. That is, for each class \(y\) and logit \(c\), there exists a nondecreasing function \(\omega_{y,c}:\mathbb{R}_+\to\mathbb{R}_+\) with \(\omega_{y,c}(0)=0\) such that, for fixed \(z\in\mathbb{R}^{p}\), two points separated by distance \(r\) in the pruned space have \(U\)-gradients that differ by at most \(\omega_{y,c}(r)\). Along the path, \(\|u(t)-u(0)\|_2=t\,r\), the modulus of continuity yields
\begin{align}
\label{eq:path-modulus}
\big\|\Delta\nabla_c(t)\big\|_2 \le \omega_{y,c}\!\big(t\,r\big).
\end{align}

Applying the Euclidean Cauchy–Schwarz inequality \(|\langle a,b\rangle|\le \|a\|_2\,\|b\|_2\) with \(a=\Delta\nabla_c(t)\) and \(b=\widehat{\Delta U}\) to the path integral \eqref{eq:R-def-z}, and invoking the modulus bound \eqref{eq:path-modulus}:
\begin{align}
\label{eq:R-CS}
R_c
&=\int_0^1 \!\big\langle \nabla_U f_\theta^{(c)}(z,u(t))-g_c(z),\,\Delta U\big\rangle\,dt \notag \\
&=\int_0^1 \!\big\langle \Delta\nabla_c(t),\,r\,\widehat{\Delta U}\big\rangle\,dt
=\; r \int_0^1 \!\big\langle \Delta\nabla_c(t),\,\widehat{\Delta U}\big\rangle\,dt \notag \\
\Rightarrow |R_c| 
&= r \left|\!\int_0^1 \!\big\langle \Delta\nabla_c(t),\,\widehat{\Delta U}\big\rangle\,dt\!\right| \notag \\
&\le\; r \int_0^1 \!\big|\!\big\langle \Delta\nabla_c(t),\,\widehat{\Delta U}\big\rangle\!\big|\, dt \notag \\
&\le\; r \int_0^1 \!\big\|\Delta\nabla_c(t)\big\|_2\,. \ \!\big\|\widehat{\Delta U}\big\|_2 dt \le\; r \int_0^1 \!\omega_{y,c}(tr)\, dt.
\end{align}

Define
\begin{equation}
\label{eq:Phi-def}
\Phi_{y,c}(r)
:= \Big(\int_0^1 \omega_{y,c}(tr)\,dt\Big)^{\!2}\, r^{2},
\end{equation}

so that
\begin{equation}
\label{eq:R-sq-bound}
R_c^2
\ \le\ \Phi_{y,c}\!\big(r\big).
\end{equation}

Assume $f_\theta$ is $\mathcal{C}^{1,\alpha}$ in the pruned space $U$, i.e., each logit's $U$-gradient is H\"older-continuous of order $\alpha\in(0,1]$ with respect to $\|\cdot\|_2$: there exist per-class constants $H_{\alpha,y,c}\ge 0$ such that, for all $z\in\mathbb{R}^{p}$, $u,u'\in\mathbb{R}^{q}$, and $c\in\{1,\dots,C\}$,
\begin{align}
\label{eq:holder-grad-perc}
\big\|\nabla_U f_\theta^{(c)}(z,u)-\nabla_U f_\theta^{(c)}(z,u')\big\|_2
\ \le\ H_{\alpha,y,c}\,\|u-u'\|_2^{\alpha}.
\end{align}

Along the path with $u=u(t)$ and $u'=u(0)$, we have $\|u(t)-u(0)\|_2=t\,r$. Applying \eqref{eq:holder-grad-perc}:
\begin{align}
\label{eq:path-holder-norm}
\big\|\Delta\nabla_c(t)\big\|_2
&=\big\|\nabla_U f_\theta^{(c)}(z,u(t))-\nabla_U f_\theta^{(c)}(z,u(0))\big\|_2
\notag \\ 
&\le H_{\alpha,y,c}\,\|u(t)-u(0)\|_2^{\alpha}
= H_{\alpha,y,c}\,(t\,r)^\alpha.
\end{align}

Comparing \eqref{eq:path-holder-norm} with \eqref{eq:path-modulus}, the H\"older assumption gives the explicit modulus
\begin{equation}
\label{eq:holder-modulus}
\ \omega_{y,c}(r)\;=\;H_{\alpha,y,c}\,r^\alpha.
\end{equation}
This $\omega_{y,c}$ is (i) nondecreasing on $\mathbb{R}_+$, (ii) anchored at zero, $\omega_{y,c}(0)=0$, and (iii) scales as: $\omega_{y,c}(\lambda r)=H_{\alpha,y,c}(\lambda r)^{\alpha}=\lambda^{\alpha}\,\omega_{y,c}(r)$ for $\lambda>0$ (Lipschitz case when $\alpha=1$).

Integrating the modulus of continuity along a segment of length $r$ gives:
\begin{align}
\int_0^1 \omega_{y,c}(tr)\,dt
&= \int_0^1 H_{\alpha,y,c}(tr)^\alpha dt
= \frac{H_{\alpha,y,c}}{\alpha+1}\,r^\alpha.
\end{align}

Substituting into \eqref{eq:Phi-def}:
\begin{align}
\Phi_{y,c}(r)
:= \left( \int_0^1 \omega_{y,c}(tr)\,dt \right)^{\!2}\, r^2
= \frac{H_{\alpha,y,c}^2}{(\alpha+1)^2}\, r^{2+2\alpha}.
\label{eq:holder-Phi-perc}
\end{align}

Combining \eqref{eq:holder-Phi-perc} with \eqref{eq:R-sq-bound} yields
\begin{align}
\label{eq:R-square-by-Phi}
R_c^2
\;\le\; \left(\frac{H_{\alpha,y,c}}{\alpha+1}\right)^{\!2} r^{2+2\alpha}
\;=\; \Phi_{y,c}(r).
\end{align}

Bounding $B_y$ from \eqref{eq:B-def}:
\begin{align}
B_y
&= \sum_{c=1}^C \mathbb{E}\!\left[(R_c-\mathbb{E}[R_c\mid Z,Y=y])^{2}\,\Big|\,Y=y\right] \notag\\[-2pt]
&\le \sum_{c=1}^C \mathbb{E}\!\left[R_c^2\,\Big|\,Y=y\right] \notag\\[-2pt]
&\le \sum_{c=1}^C \mathbb{E}\!\left[\Phi_{y,c}\!\big(r\big)\,\Big|\,Y=y\right] \notag\\[-2pt]
&= \sum_{c=1}^C \mathbb{E}\!\left[\frac{H_{\alpha,y,c}^2}{(\alpha+1)^2}\, r^{2+2\alpha}\,\Big|\,Y=y\right] \notag\\[-2pt]
&= \left(\!\sum_{c=1}^C \frac{H_{\alpha,y,c}^{2}}{(\alpha+1)^{2}}\!\right)
\mathbb{E}\!\left[\|U{-}\mu_y(Z)\|_2^{\,2+2\alpha}\,\Big|\,Y{=}y\right].
\label{eq:By-bound-explicit-perc}
\end{align}

If, moreover,
\begin{align*}
\left(\!\sum_{c=1}^C \frac{H_{\alpha,y,c}^{2}}{(\alpha+1)^{2}}\!\right)
\mathbb{E}\!\left[\|U{-}\mu_y(Z)\|_2^{\,2+2\alpha}\,\middle|\,Y{=}y\right]
\!\le\! L_y,
\end{align*}
then substituting this bound on $B_y$ into \eqref{eq:psi-bound} yields \eqref{eq:psi-holder-perc}, completing the proof.
\end{proof}

\section{Proof of Proposition~2}
\label{app:prop-affine}

\begin{proof}

Under the affine assumption, for fixed \(z\), the $c$-th logit $f_\theta^{(c)}$ is affine along the segment $u(t)$: there exist \(\varkappa_c(z)\in\mathbb{R}\) and \(g_c(z)\in\mathbb{R}^{q}\) such that \(f_\theta^{(c)}(z,u(t))=\varkappa_c(z)+g_c(z)^\top u(t)\) for all $t\in[0,1]$. Taking the $U$-gradient:
\begin{align}
    \nabla_U f_\theta^{(c)}\big(z,u(t)\big) = \nabla_u [\varkappa_c(z) + g_c(z)^\top u(t)]= g_c(z).
\end{align}

Since $\nabla_U f_\theta^{(c)}(z,u(t)) \equiv g_c(z)$, the integrand in \eqref{eq:R-def-z} vanishes, yielding \(R_c=0\) and \(B_y=0\) by \eqref{eq:B-def}.

For random variables, the affine structure yields $f_\theta^{(c)}(Z,U) = \varkappa_c(Z) + g_c(Z)^\top U$ where $\varkappa_c(Z)$ and $g_c(Z)$ are deterministic given $Z$. The conditional variance is:
\begin{align}
&\text{Var}(f_\theta^{(c)}(Z,U)|Z,Y=y) 
\notag\\
&= \text{Var}(\varkappa_c(Z) + g_c(Z)^\top U|Z,Y=y)  \notag\\
&= \text{Var}(g_c(Z)^\top U|Z,Y=y) = g_c(Z)^\top \Sigma_y^-(Z) g_c(Z). 
\end{align}

Applying $\mathbb{E}[\cdot|Y=y]$ and summing over $c\in[C]$:
\begin{align}
&\sum_{c=1}^C \mathbb{E}[\text{Var}(f_\theta^{(c)}(Z,U)|Z,Y=y)|Y=y] 
\notag \\
&= \sum_{c=1}^C \mathbb{E}[g_c(Z)^\top \Sigma_y^-(Z) g_c(Z)|Y=y] \notag\\
&= \mathbb{E}\left[\sum_{c=1}^C g_c(Z)^\top \Sigma_y^-(Z) g_c(Z)\,\Big|\,Y=y\right] \notag\\
&= L_y \quad \text{by } \eqref{eq:L-def}.
\end{align}

By \eqref{eq:step1-varsum}, taking the supremum over $y\in[C]$ yields \eqref{eq:psi-affine-equality}.

\end{proof}

\begin{algorithm*}[p]
\caption{Empirical estimates of the irreducibility bound components}
\label{alg:validation_single}
\begin{small}
\SetKwInOut{Input}{Input}
\Input{Global correlation cutoff $\tau\in(0,1)$ defining retained index set $\widehat{\mathcal{Q}}_\tau\subseteq[d]$ and pruned indices $\widehat{\mathcal{Q}}_\tau^{-} := [d]\setminus \widehat{\mathcal{Q}}_\tau$. Sample matrices $\{X^{(y)}\}_{y=1}^C$ for each class $y\in[C]$, where $X^{(y)} \in \mathbb{R}^{n_y\times d}$. Differentiable classifier $f_\theta:\mathbb{R}^d\to\mathbb{R}^C$ (logits); Monte Carlo (MC) draws $n_U\in\mathbb{N}$; curvature probe grid $\mathcal{T}\subset(0,1]$; H\"older exponent $\alpha\in(0,1]$.}
\BlankLine
\For{$y \in \{1,\dots,C\}$}{
  \textbf{Split kept/pruned.}
  \quad $Z := X^{(y)}_{:,\,\widehat{\mathcal{Q}}_\tau} \in \mathbb{R}^{n_y\times p}$, \
        $U := X^{(y)}_{:,\,\widehat{\mathcal{Q}}_\tau^{-}} \in \mathbb{R}^{n_y\times q}$,
        where $p=|\widehat{\mathcal{Q}}_\tau|$, $q=|\widehat{\mathcal{Q}}_\tau^{-}|$, and $d=p+q$\;

  \textbf{Fit linear \(\mathbf{U\mid Z}\) by Ordinary Least Squares (OLS) with intercept.}
  \quad $Z_1 := [\,Z\ \ \mathbf{1}\,] \in \mathbb{R}^{n_y\times(p+1)}$.
  \quad Solve $\begin{bmatrix}\widehat{A}^\top \\ \widehat{b}^\top\end{bmatrix} := Z_1^{+} U$ \ (or $(Z_1^\top Z_1)^{-1} Z_1^\top U$ if full rank),
  so $\widehat{A}\in\mathbb{R}^{q\times p}$, $\widehat{b}\in\mathbb{R}^{q}$; here $Z_1^{+}$ denotes the \emph{Moore–Penrose pseudoinverse} of $Z_1$\;

  \textbf{Anchor \& residual covariance.}
  \quad \text{Anchor:} $\widehat{\mu}_y(Z) := Z \widehat{A}^\top + \mathbf{1}\,\widehat{b}^\top \in \mathbb{R}^{n_y\times q}$; \ \text{Residual:}
        $\Delta U^{(y)} := U - \widehat{\mu}_y(Z) \in \mathbb{R}^{n_y\times q}$; \ \text{Covariance:}
        \[\Sigma_y^{-} := \dfrac{1}{n_y-1}\,(\Delta U^{(y)})^\top(\Delta U^{(y)}) \in \mathbb{R}^{q\times q}.\]

  \textbf{Gradients at the anchor (pruned indices).}
  \quad For each sample $i$, define $x_i^{\mathrm{anc}} := \big(Z_i,\ \widehat{\mu}_y(Z)_i\big)\in\mathbb{R}^{p+q}$.
  For $c=1,\dots,C$, let $J_i^{(c)} := \nabla_{X} f_\theta^{(c)}(x_i^{\mathrm{anc}}) \in \mathbb{R}^{1 \times d}$.
  Form the per-class Jacobian matrix
  \[
  J^{(c)} := \begin{bmatrix}
  \nabla_X f_\theta^{(c)}(x_1^{\mathrm{anc}})\\[-2pt]
  \vdots\\[-2pt]
  \nabla_X f_\theta^{(c)}(x_{n_y}^{\mathrm{anc}})
  \end{bmatrix} \in \mathbb{R}^{n_y\times d},\quad
  G^{(c)} := J^{(c)}_{:,\,Q_{\mathrm{prune}}} \in \mathbb{R}^{n_y\times q},
  \]
  and define $g_i^{(c)} := \big(G^{(c)}_{i,\cdot}\big)^\top \in \mathbb{R}^{q}$ (column)\;

  \textbf{Signal term ($\mathbf{\widehat{L}}_{\mathbf{y}}$).}
  \quad Stack for sample $i$: $G_i := \begin{bmatrix} (g_i^{(1)})^\top \\ \vdots \\ (g_i^{(C)})^\top \end{bmatrix} \in \mathbb{R}^{C\times q}$.
  \[
  \ell_i := \sum_{c=1}^C \underbrace{(g_i^{(c)})^\top}_{\mathbb{R}^{1 \times q}} \underbrace{\Sigma_y^{-}}_{\mathbb{R}^{q \times q}} \underbrace{g_i^{(c)}}_{\mathbb{R}^{q \times 1}}
  \;=\; \operatorname{tr}\!\Big(\underbrace{G_i}_{\mathbb{R}^{C \times q}} \underbrace{\Sigma_y^{-}}_{\mathbb{R}^{q \times q}} \underbrace{G_i^\top}_{\mathbb{R}^{q \times C}}\Big),\qquad
  \widehat{L}_y := \frac{1}{n_y}\sum_{i=1}^{n_y} \ell_i .
  \]

  \textbf{MC $\mathrm{\mathbf{LHS}}_{\mathbf{y}}$ (conditional variance over $\mathbf{U\mid Z}$).}
  \quad For each $i$, draw $U_{i,m}\sim \mathcal{N}\!\big(\widehat{\mu}_y(Z)_i,\Sigma_y^{-}\big)$ for $m=1,\dots,n_U$; let
  \[
  \bar f_{i,c} := \frac{1}{n_U}\sum_{m=1}^{n_U} f_\theta^{(c)}(Z_i,U_{i,m}), \quad
  \widehat{\mathrm{Var}}\!\left(f_\theta^{(c)} \mid Z_i\right)
  := \frac{1}{n_U-1}\sum_{m=1}^{n_U}\!\Big(f_\theta^{(c)}(Z_i,U_{i,m}) - \bar f_{i,c}\Big)^2 .
  \]
  \[
  \mathrm{LHS}_y := \frac{1}{n_y}\sum_{i=1}^{n_y} \sum_{c=1}^C \widehat{\mathrm{Var}}\!\left(f_\theta^{(c)} \mid Z_i\right).
  \]

  \textbf{Curvature penalty ($\mathbf{\widehat{B}}_{\mathbf{y}}$) via H\"older\textendash$\mathbf{\alpha}$ gradient modulus.}
  \quad For each sample $i$, set $\Delta U_i^{(y)} := U_i - \widehat{\mu}_y(Z)_i$, $r_i := \|\Delta U_i^{(y)}\|_2$, and $\tilde{\mathcal{I}} := \{\,i : r_i>0\,\}$. Define the path $U_i(t) := \widehat{\mu}_y(Z)_i + t\,\Delta U_i^{(y)}$ and the unit direction $\widehat{\Delta U}_i := \Delta U_i^{(y)}/r_i$ for $i\in\tilde{\mathcal{I}}$. For $t\in\mathcal{T}$ and each class $c$, define the \emph{directional drift}
  \[
    \mathrm{drift}_{i,c}^{\parallel}(t)
    \;:=\;
    \big|\ \langle\, \nabla_U f_\theta^{(c)}(Z_i,U_i(t)) - \nabla_U f_\theta^{(c)}(Z_i,\widehat{\mu}_y(Z)_i),\ \widehat{\Delta U}_i \rangle\ \big| .
  \]
  Estimate per-class H\"older constants:
  \[
    \rho_{i,c}(t) \;:=\; \frac{\mathrm{drift}_{i,c}^{\parallel}(t)}{t^{\alpha}\, r_i^{\alpha}},\qquad
    H_{\alpha,y,c}\ :=\ \sup_{i\in\tilde{\mathcal{I}},\,t\in\mathcal{T}}\,\rho_{i,c}(t).
  \]
  With $\omega_{y,c}(r)\approx H_{\alpha,y,c}\,r^{\alpha}$, the H\"older modulus of \eqref{eq:holder-modulus}, define
  \[
    \widehat{B}_y \;:=\;
    \frac{\sum_{c=1}^C H_{\alpha,y,c}^{2}}{(\alpha+1)^{2}}\cdot
    \frac{1}{n_y}\sum_{i=1}^{n_y} r_i^{\,2+2\alpha}.
  \]

  \textbf{Per-class $\mathrm{\mathbf{RHS}}_{\mathbf{y}}$.}
  \quad $\mathrm{RHS}_y := \big(\sqrt{\widehat{L}_y} - \sqrt{\widehat{B}_y}\big)^{2}$.
}
\textbf{Supremum over classes (irreducibility bound verification).}
\hspace{2em}
\(\displaystyle
\widehat{\Psi}_\theta(\tau) := \mathrm{LHS}_{\sup} = \max_{y\in[C]} \text{LHS}_y,\quad
\mathrm{RHS}_{\sup} := \max_{y\in[C]} \mathrm{RHS}_y,\qquad
\boxed{\ \mathrm{LHS}_{\sup} \ \ge\ \mathrm{RHS}_{\sup}\ }.
\)

\KwRet{margin $:= \mathrm{LHS}_{\sup}-\mathrm{RHS}_{\sup}$}
\end{small}
\end{algorithm*}

\section{Pseudocode for the Experimental Evaluation of Corollary~1}
\label{app:bound_evaluation}
The computational procedure for validating Corollary~1 with experimental dataset is detailed in Algorithm~\ref{alg:validation_single}.

\makeatletter
\def\@IEEEBIOskipN{1\baselineskip}
\makeatother
\bibliographystyle{IEEEtran}
\bibliography{references}

\begin{IEEEbiography}[{\includegraphics[width=1in,height=1.25in]{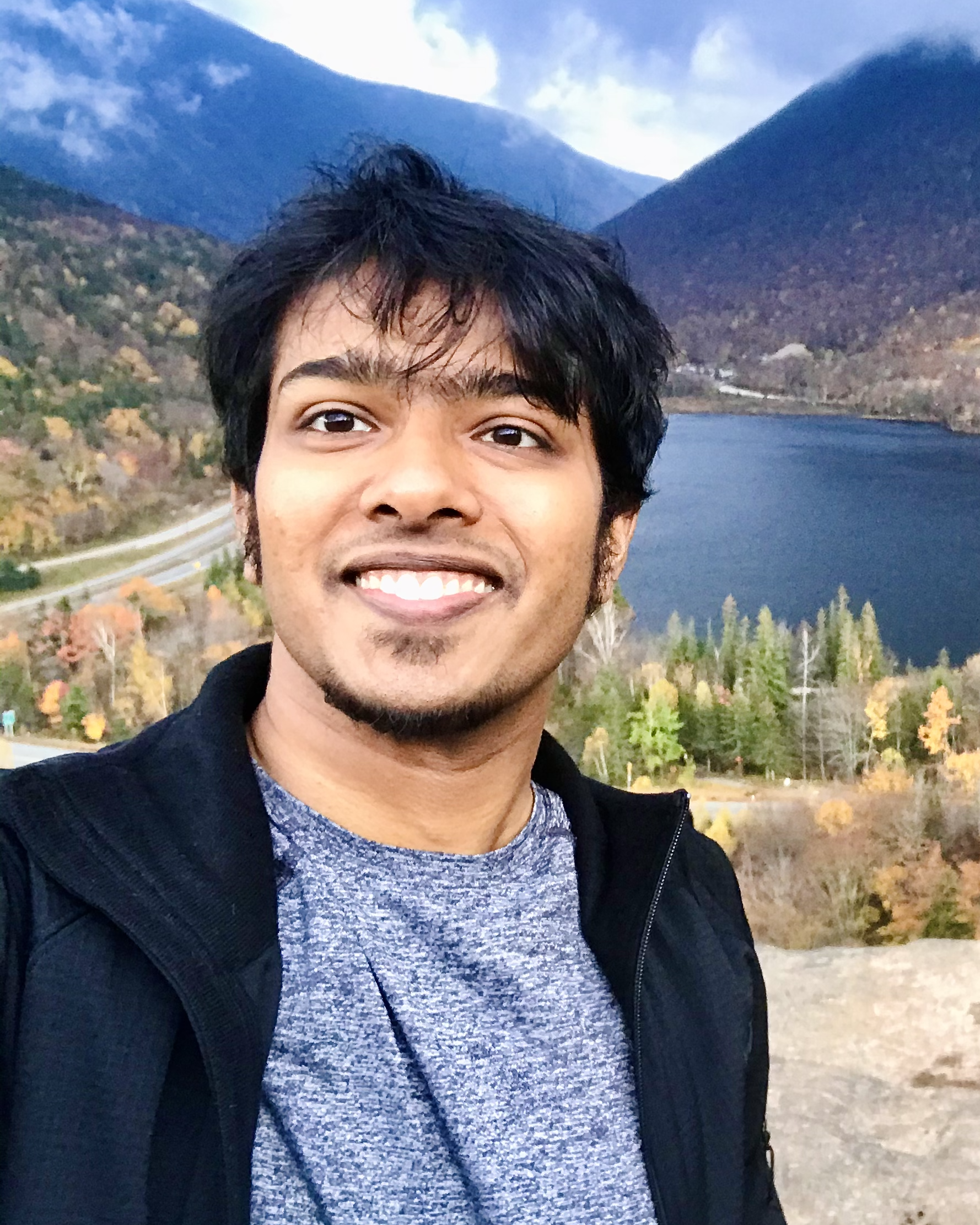}}]{Abdullah Al Bashit}
received the B.Sc. degree in Electrical \& Electronic Engineering from Rajshahi University of Engineering \& Technology (RUET), Rajshahi, Bangladesh, in \(2011\), the M.Sc. degree in Engineering from Texas State University, San Marcos, TX, USA, in \(2019\), and the Ph.D. degree in Electrical Engineering from Northeastern University, Boston, MA, USA, in \(2024\). He is currently a Postdoctoral Research Associate with the Department of Bioengineering, Northeastern University, Boston, MA, USA. His research interests span statistical methods and machine learning for biomedical signal and image analysis. His awards and recognitions include, but are not limited to, the American Crystallographic Association (ACA) Etter Student Lecture Award (\(2021\), \(2023\)), and Texas State University's Director's List (\(2017\)--\(2019\)).
\end{IEEEbiography}

\begin{IEEEbiography}[{\includegraphics[width=1in,height=1.25in]{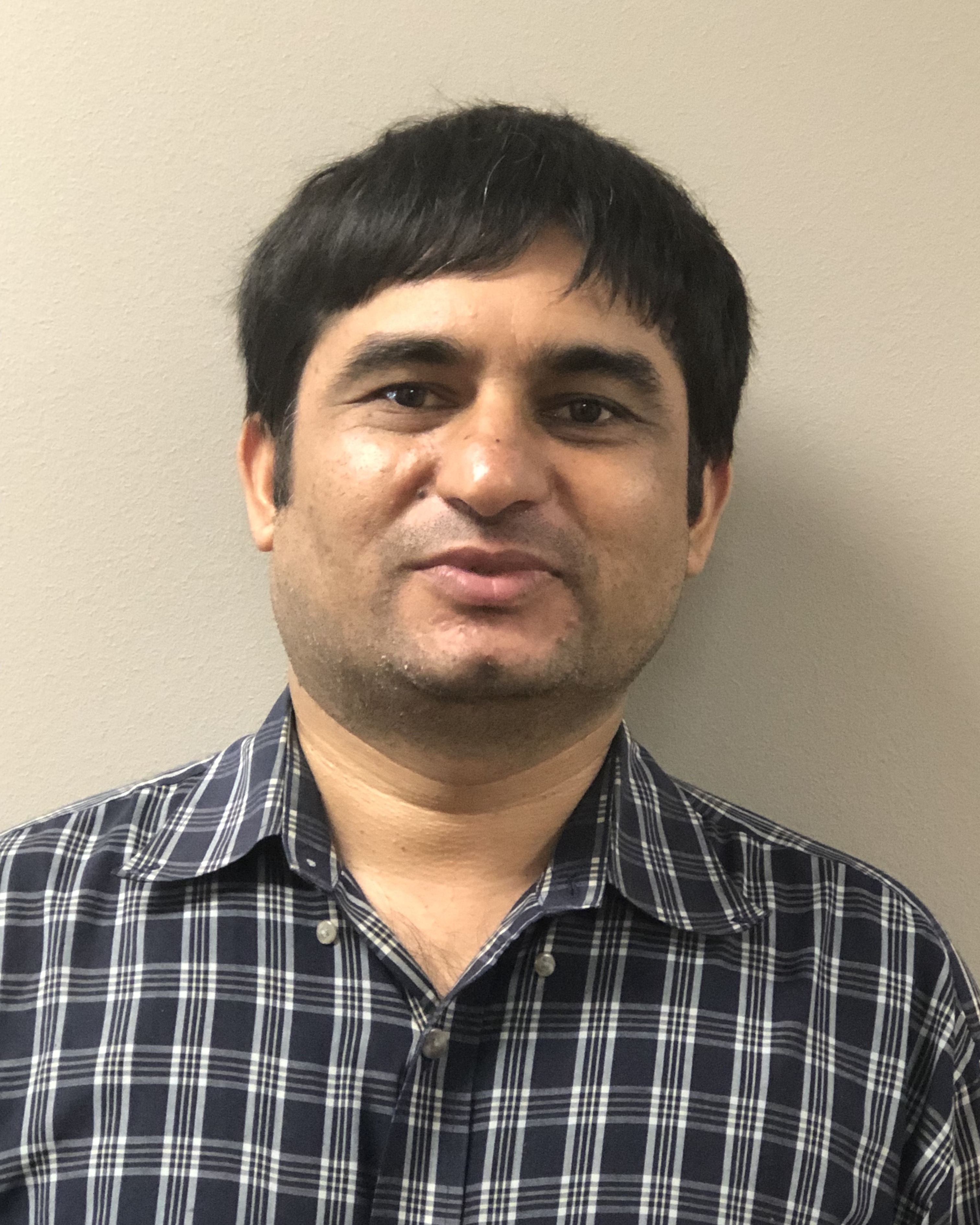}}]{Prakash Nepal}
received the B.Sc. and M.Sc. degrees in Physics from Tribhuvan University, Nepal, in \(2003\) and \(2006\), respectively, and the Ph.D. degree in Physics from the University of Wisconsin–Milwaukee, Milwaukee, WI, USA, in \(2017\). He completed postdoctoral research at the University of Wisconsin–Milwaukee, Milwaukee, WI, USA, and at Louisiana State University, Baton Rouge, LA, USA, and joined Northeastern University, Boston, MA, USA, in December 2020, where he is a Research Scientist in Lee Makowski’s laboratory. His research focuses on X-ray scattering and Imaging, Biophysics, and the intersection of Theory and Computation. He has authored multiple publications in diverse fields, with work appearing in journals such as Physical Review, Classical Quantum Gravity, and Applied Crystallography.
His awards and recognitions include BioXFEL (Biology with X-ray Free Electron Lasers) honors, namely the MacGyver Award for a grant-proposal competition (\(2015\)) and the Best Poster Award at the BioXFEL International Conference (\(2017\)), and multiple Research Excellence Awards at the University of Wisconsin–Milwaukee (\(2012\)--\(2017\)).
\end{IEEEbiography}

\begin{IEEEbiography}[{\includegraphics[width=1in,height=1.25in,clip,keepaspectratio]{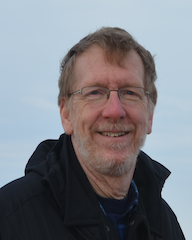}}]{Lee Makowski}
is Professor of Bioengineering and of Chemistry and Chemical Biology at Northeastern University, Boston, MA, USA. He holds a B.S. in Physics from Brown University, Providence, RI, USA, and a Ph.D. in Electrical Engineering from the Massachusetts Institute of Technology (MIT), Cambridge, MA, USA. In \(2010\), he moved from Argonne National Laboratory, Argonne, IL, USA, where he served as Biosciences Division Director and Senior Scientist, to Northeastern University, Boston, MA, USA, where he accepted a joint appointment in Electrical and Computer Engineering and in Chemistry and Chemical Biology. He founded the Department of Bioengineering at Northeastern University, Boston, MA, USA, in January \(2014\) and was subsequently named Inaugural Chair, a position he held until January \(2024\). His career has been highlighted by the use of state-of-the-art signal processing methods to study macromolecular systems not amenable to standard approaches, with recent emphasis on the molecular basis of neurodegenerative processes, including Alzheimer’s disease. He has built partnerships with organizations in Ghana over the past \(8\) years. He is a co-founder of the BioInnovation Center at Academic City University, Accra, Ghana, which develops programs to foster entrepreneurial activity in the medical device sector.
\end{IEEEbiography}

\end{document}